\documentclass[aps,prx,superscriptaddress,longbibliography, twocolumn]{revtex4-1} 
\pdfoutput=1
\usepackage[version=3]{mhchem} % Formula subscripts using \ce{}
\usepackage[dvips]{epsfig}
\usepackage{graphicx}  % needed for figures
\usepackage{dcolumn}   % needed for some tables
\usepackage{bm}        % for math
\usepackage{amssymb}   % for math
\usepackage{color}
\usepackage{float}

\begin{document}

\author{Behzad Khanaliloo}
\affiliation{Institute for Quantum Science and Technology, University of Calgary, Calgary, AB, T2N 1N3, Canada}
\affiliation{National Institute for Nanotechnology, Edmonton, AB, T6G 2M9, Canada}

\author{Harishankar Jayakumar}
\affiliation{Institute for Quantum Science and Technology, University of Calgary, Calgary, AB, T2N 1N3, Canada}

\author{Aaron C. Hryciw}
\affiliation{National Institute for Nanotechnology, Edmonton, AB, T6G 2M9, Canada}
\affiliation{University of Alberta nanoFAB, Edmonton, AB, T6G 2R3, Canada}

\author{David P. Lake}
\affiliation{Institute for Quantum Science and Technology, University of Calgary, Calgary, AB, T2N 1N3, Canada}

\author{Hamidreza Kaviani}
\affiliation{Institute for Quantum Science and Technology, University of Calgary, Calgary, AB, T2N 1N3, Canada}

\author{Paul E. Barclay}
\email{pbarclay@ucalgary.ca}
\affiliation{Institute for Quantum Science and Technology, University of Calgary, Calgary, AB, T2N 1N3, Canada}
\affiliation{National Institute for Nanotechnology, Edmonton, AB, T6G 2M9, Canada}

\title{Single crystal diamond nanobeam waveguide optomechanics}

\begin{abstract}
 
Optomechanical devices sensitively transduce and actuate motion of nanomechanical structures using light. Single--crystal diamond promises to improve the performance of optomechanical devices, while also providing opportunities to interface nanomechanics with diamond color center spins and related quantum technologies. Here we demonstrate dissipative waveguide--optomechanical coupling exceeding 35 GHz/nm to diamond nanobeams supporting both optical waveguide modes and mechanical resonances, and use this optomechanical coupling to measure nanobeam displacement with a sensitivity of $9.5$ fm/$\sqrt{\text{Hz}}$ and optical bandwidth $>150$nm. The nanobeams are fabricated from bulk optical grade single--crystal diamond using a scalable undercut etching process, and  support mechanical resonances with quality factor  $2.5 \times 10^5$ at room temperature, and $7.2 \times 10^5$ in cryogenic conditions (5K).  Mechanical self--oscillations, resulting from interplay between photothermal and optomechanical effects, are observed with amplitude exceeding 200 nm for sub-$\mu$W absorbed optical  power, demonstrating the potential for optomechanical excitation and manipulation of diamond nanomechanical structures.

\end{abstract}

\maketitle

%%% INTRO %%%

% Optomechanical Nanomechanical devices: good
% Diamond: better.
% Motivate why

% Large dynamics range

Nanophotonic optomechanical devices \cite{ref:li2008hof, ref:eichenfield2009oc, ref:gavartin2011oct, ref:sun2012fdc} allow chip--based optical control of nanomechanical resonances with a precision reaching the standard quantum limit \cite{ref:arcizet2006hso, ref:schliesser2009rsc, ref:chan2011lcn, ref:cohen2013ocn}, enabling  tests of quantum nanomechanics \cite{ref:teufel2009nmm, ref:chan2011lcn, ref:verhagen2012qcc, ref:safavinaeini2012oqm}, as well as technologies for sensing \cite{ref:anetsberger2009nco, ref:srinivasan2011oti, ref:sun2012fdc, ref:krause2012ahm} and information processing \cite{ref:bagheri2011dmn, ref:hill2012cow, ref:bochmann2013ncb}.  Single crystal diamond  devices are particularly exciting for these applications.  In addition to possessing desirable mechanical and optical properties, diamond hosts color centers \cite{ref:gruber1997sco}  whose highly coherent electronic and nuclear spins are promising for quantum technologies. Recently, piezoelectric--actuated nanomechanical manipulation of electronic spins in diamond color centers \cite{ref:macquarrie2013msc, ref:ovartchaiyapong2014dsc, ref:teissier2014scn} and quantum dots \cite{ref:yeo2013smc, ref:montinaro2014qdo} have illuminated the possibility of using diamond nanomechanical devices \cite{ref:sekaric2002nrs, ref:ovartchaiyapong2012hqf, ref:tao2013scd, ref:rath2013doc, ref:burek2013nrs} to mediate phonon--spin interactions at the quantum level. Adding optomechanical functionality to these devices will allow efficient coherent optical manipulation and readout of the nanomechanical stuctures, and provide a path towards implementing an efficient interface between optical, mechanical and electronic quantum systems. Such an interface would play a crucial role in many proposed studies, including normal mode cooling \cite{ref:wilson2004lcn, ref:kepesidis2013pcl}, quantum nonlinear optomechanics \cite{ref:ramos2013nqo}, and spin--squeezing \cite{ref:bennett2013pis}.

\begin{figure*}[tbh]
\begin{center}
\epsfig{figure=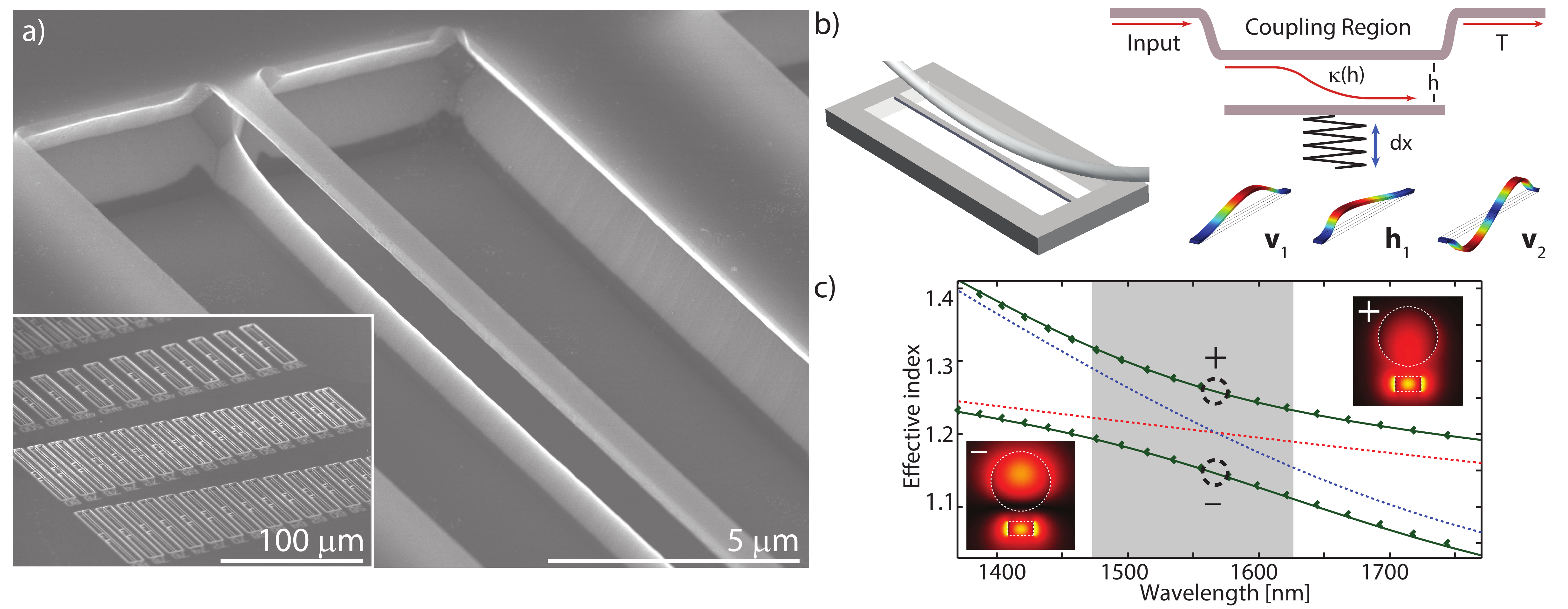, width=1\linewidth}
 \caption{Diamond nanobeam waveguide optomechanical system. (a), Scanning electron micrograph (SEM) images of a single crystal diamond nanobeam--waveguide. Dark high--contrast regions are due to variations in thickness of titanium deposited for  imaging purposes. (b), Schematic of fiber taper evanescent waveguide coupling geometry and illustration of the waveguide--optomechanical coupling process. Nanobeam mechanical resonances, whose typical displacement profiles are shown, modulate the distance between the input waveguide, changing $\kappa(h)$. (c), Dispersion of  $n_{\text{eff}}$ of the waveguide modes of the fiber taper (diameter $1.1\,\mu$m) and diamond nanobeam ($w\times d = 460 \times 250\,\text{nm}^2$) when they are uncoupled (dashed lines, red: fiber taper, blue: nanobeam) and coupled with $h = 300$ nm (solid lines). Shaded region indicates the laser scan range used in the experiments. Insets: mode profiles ($x-y$ plane, dominant electric field component $|E_y|$)  of the symmetric and antisymmetric coupled modes. }
\label{fig:schematic}
\end{center}
\end{figure*}

Here we demonstrate a diamond waveguide optomechanical system  providing sensitive and broadband readout of ultrahigh mechanical quality factor ($Q_m$) resonances which can be excited into large amplitude optomechanical self-oscillations, revealing the nonlinear nanomechanical  properties of the nanobeams.  Dissipative optomechanical coupling in this system relies on efficient phase--matched evanescent coupling between a fiber taper waveguide and optical modes of ultrahigh--$Q_m$ diamond nanobeam resonators. The nanobeams are fabricated from single--crystal diamond chips using oxygen plasma undercut etching. They support mechanical resonances whose motion is transduced by the optical fiber taper--nanobeam coupling, resulting in an optomechanical interaction characterized by an effective dissipative optomechanical coupling coefficient of $35 \,\text{GHz/nm}$, and a displacement sensitivity of $9.5\, \text{fm}/\sqrt{\text{Hz}}$.  In combination with low nanobeam mechanical dissipation ($Q_m > 7.2\times 10^5$), interplay between pN  photothermal forces and optomechanical coupling is shown to excite nanomechanical self--oscillations with amplitude $> 200\,\text{nm}$, which renormalize the nanobeam mechanical frequency and provide a measure of the nanobeam's internal stress and buckling amplitude. The excellent displacement sensitivity, broadband coupling, and ultrahigh--$Q_m$ demonstrated here will enable diamond nanobeam waveguide optomechanical implementations of sensors \cite{ref:tao2013scd, ref:verhagen2012qcc} and quantum nanomechanical systems \cite{ref:schliesser2009rsc, ref:chan2011lcn}, while the ability to optomechanically excite nanomechanical resonances shows the potential for optical manipulation of local stress fields and accompanying phonon--spin coupling in diamond color centers \cite{ref:rabl2010qst, ref:arcizet2011snv, ref:hong2012cmc, ref:macquarrie2013msc, ref:ovartchaiyapong2014dsc, ref:teissier2014scn}, and realization of hybrid photon--spin--phonon systems \cite{ref:ramos2013nqo}.

\section{ Waveguide optomechanics}

Experiments in diamond nanophotonics and nanomechanics have recently been advanced by the availability of high--quality diamond chips grown using chemical vapor deposition.  While  optomechanical devices can be fabricated from polycrystalline diamond films \cite{ref:rath2013doc}, single crystal diamond thin films which retain the desirable combination of optical, mechanical and quantum electronic properties introduced above are not commercially available, and must be manufactured using ion-implantation \cite{ref:fairchild2008fus, ref:magyar2011ftl,ref:patton2012ops} or wafer-bonding \cite{ref:faraon2011rez, ref:ovartchaiyapong2012hqf, ref:tao2013scd, ref:bayn2014ftn} techniques. Fabrication of devices directly from bulk single crystal diamond chips is desirable, and until now has relied on either ion beam milling \cite{ref:riedrich2012otd}, or less damaging and more scalable inductively coupled plasma reactive ion (ICPRIE) Faraday cage angled--etching \cite{ref:burek2013nrs, ref:burek2014hqf}.   In this work we demonstrate an ICPRIE process which does not require a Faraday cage, and instead utilizes diamond undercut etching to fabricate nanobeams, such as those shown in Fig.\ \ref{fig:schematic}(a), from bulk single crystal diamond.  This process, summarized in Fig.\ \ref{fig:nanofab} and discussed in more detail in Appendix \ref{sec:fab}, is inspired by earlier bulk silicon nanofabrication \cite{ref:shaw1994scr} and relies on an oxygen based ICPRIE process operated with zero--RF power, high--ICP power and an elevated sample temperature ($250\,^\circ$C) to etch diamond  quasi--isotropically along diamond crystal planes. The process is uniform over the diamond chip area and is fully compatible with standard commercial etching tools. 

Using this diamond undercutting process, nanobeams were fabricated with widths $w = 300-540\,\text{nm}$ and $750\,\text{nm}$ and lengths $L = 50 - 80\,\mu\text{m}$ on the same chip. Nanobeam thickness $d$ was adjusted by controlling the etching times together with the nanobeam width $w$. For the etch parameters used here, the nanobeam thickness was $d\sim250 - 350\,\text{nm}$, depending on $w$. Owing to the crystal--plane sensitive nature of the undercut etch, for the etch parameters used here the narrow nanobeams have a flat bottom surface, while the $w=750\,\text{nm}$ nanobeams have a triangular bottom surface mimicking the ridge visible in Fig.\ \ref{fig:schematic}(a).  These bottom surfaces can be flattened by longer undercut etching.

\begin{figure*}[t]
\begin{center}
\epsfig{figure=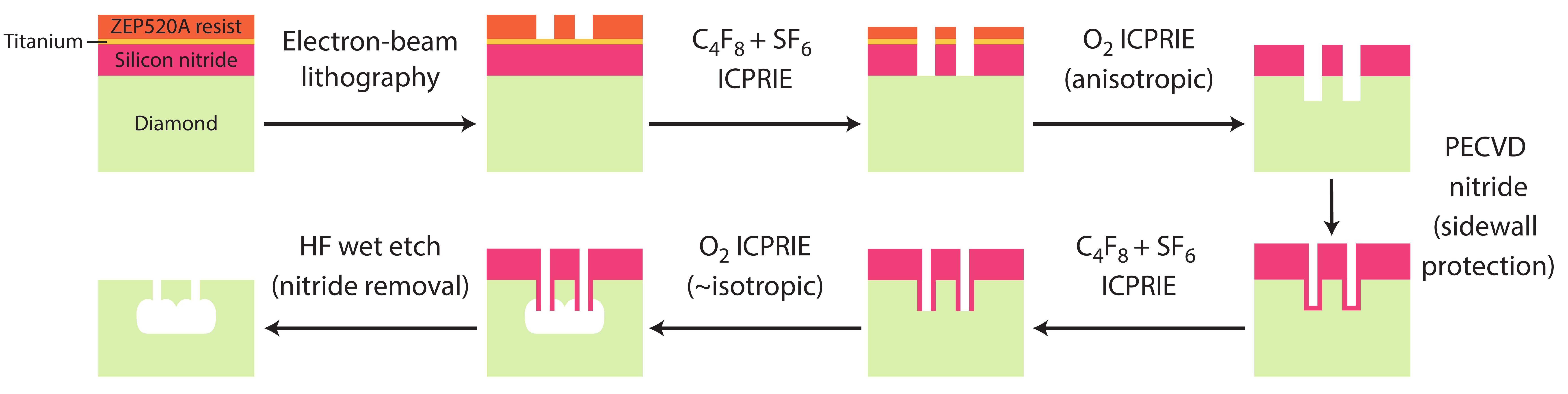, width=1\linewidth}
 \caption{Process flow for creating diamond nanobeams using quasi--isotropic reactive ion undercut etching.}
\label{fig:nanofab}
\end{center}
\end{figure*}

\begin{figure}[t]
\begin{center}
\epsfig{figure=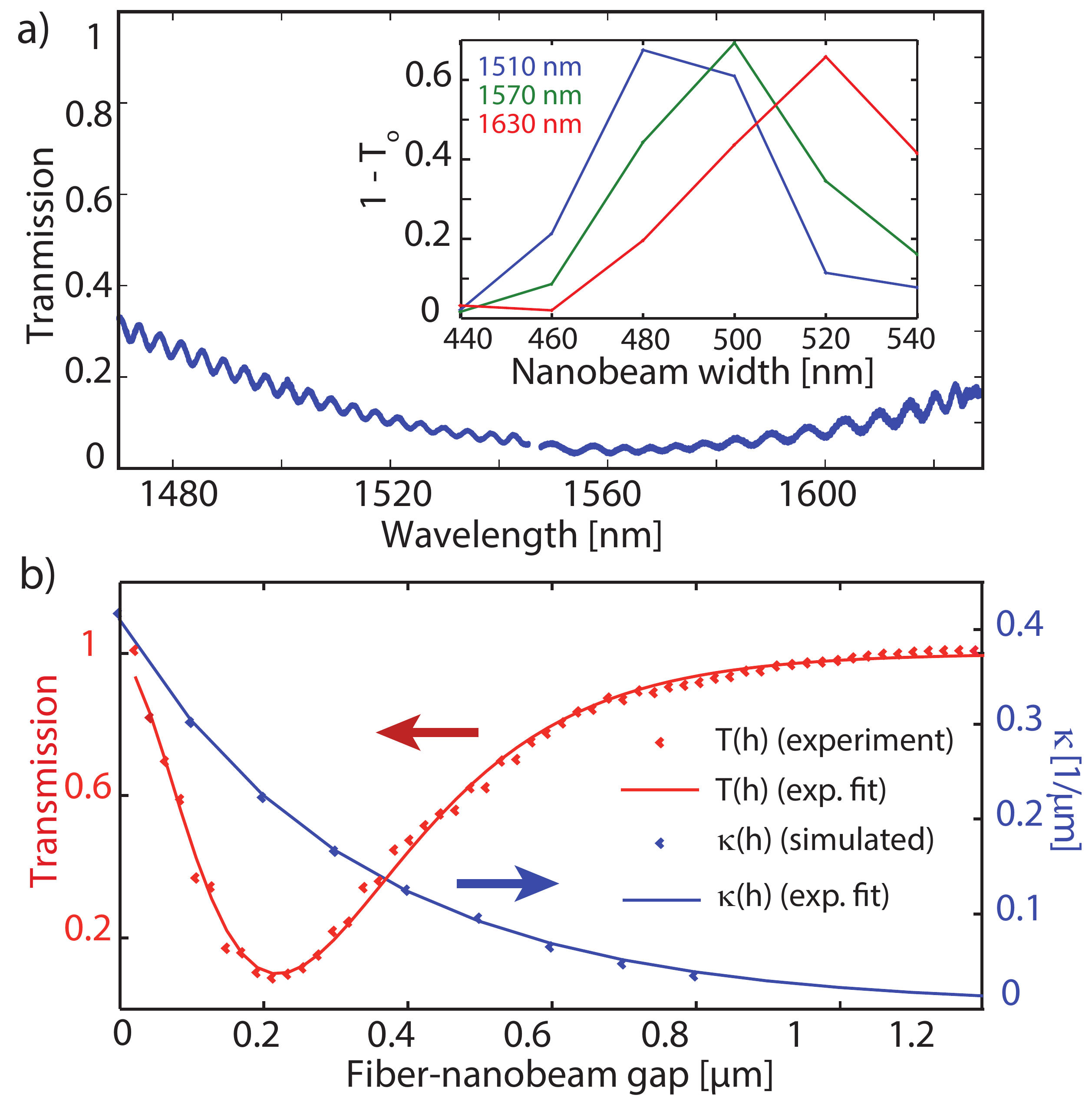, width=1\linewidth}
 \caption{Efficient evanescent waveguide coupling. (a), Fiber taper transmission  $\overline{T}(\lambda)$ vs.\ wavelength for $h = 200\,\text{nm}$. Inset: variation of  $\overline{T}$ with nanobeam width $w$ and $\lambda$ for approximately constant $h$. (b), Dependence of fiber taper transmission on fiber height. Experimentally measured values for $\overline{T}_o$ (red points) are fit with the model described by Eq.\ \eqref{eq:transmission} (red line). The corresponding $\kappa(h)$ extracted from the fit (blue line) and predicted from simulation (blue points) are also shown.}
\label{fig:waveguide_coupling}
\end{center}
\end{figure}

\begin{figure*}
\begin{center}
\epsfig{figure=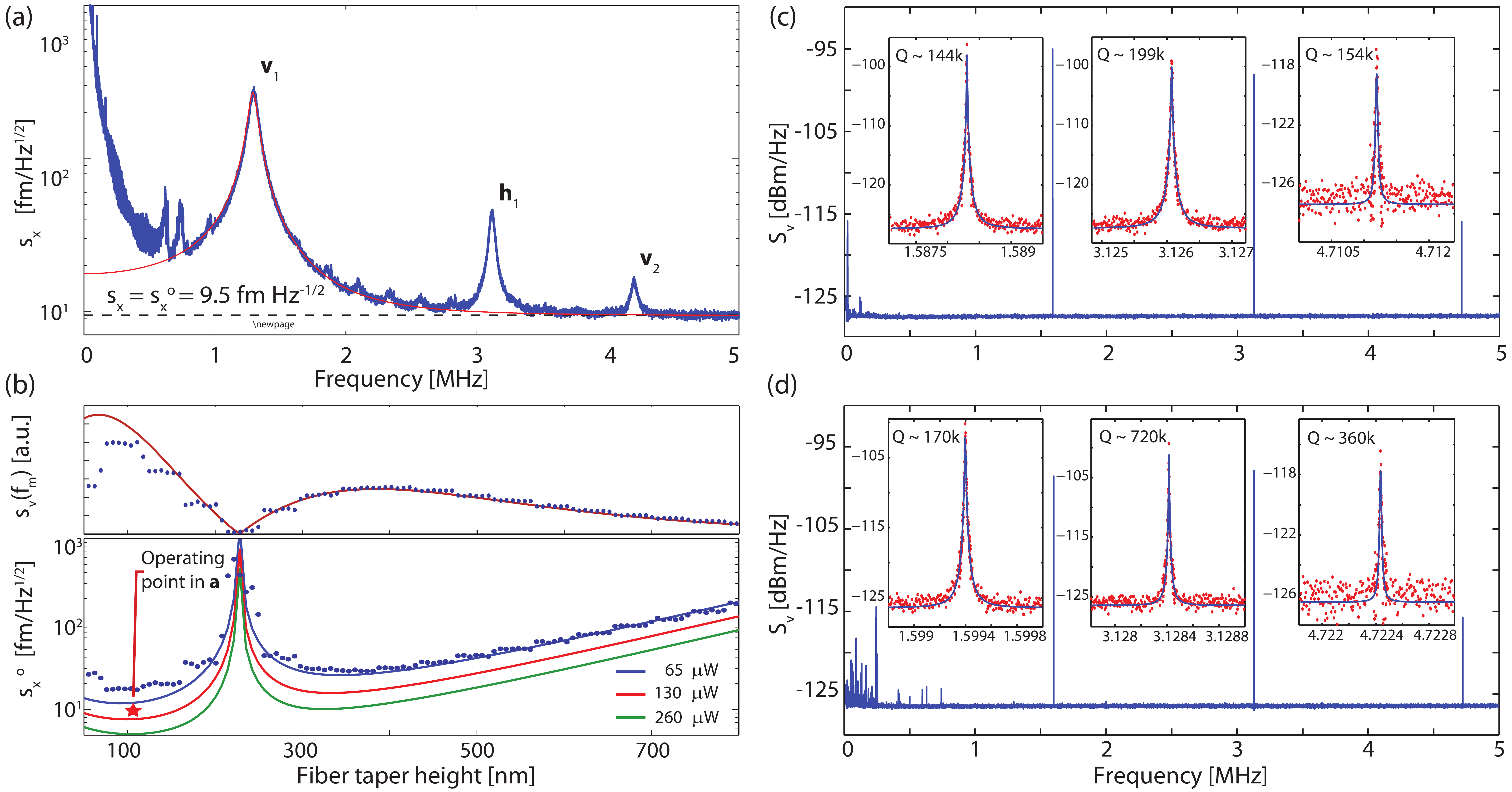, width=1\linewidth}
 \caption{Nanomechanical and optomechanical properties. (a), Measured $s_x(f)$ when the fiber taper is coupled to a nanobeam ($L\times w \times d = 60 \times 0.46 \times 0.25\,\mu\text{m}^3$) in ambient conditions, with $P_d \sim 100\,\mu$W. The vertical axis and noise--floor (dashed line) of $s_x^o = 9.5$ fm/$\sqrt{\text{Hz}}$ were calibrated from the fit to the $f= 1.3$ MHz mechanical resonance thermomechanical displacement spectrum (red solid line). (b), Top: $S_v(f_1)^{1/2}$ observed (points) and fit with a function $\propto |d\overline{T}/dh|$ (solid line). Bottom: displacement sensitivity $s_x^o(h)$ of the fundamental out--of--plane resonance from measurement (blue points) and predicted from the measured $T(h)$ for varying input power (solid lines).  The starred point corresponds the sensitivity and operating condition of the measurement in (a). Measured $S_v(f)$ for a nanobeam with $L \times w\times d = 60 \times 0.75\times 0.3\,\mu\text{m}^3$ (c), in vacuum (d), at 5K. Insets: fits to observed mechanical resonances, and corresponding $Q_m$.}
\label{fig:optomechanics}
\end{center}
\end{figure*}

The fabricated nanobeams support mechanical resonances, the lowest order of which are illustrated in Fig.\ \ref{fig:schematic}(b), with frequencies and effective mass in the $\text{MHz}$  and pg range, respectively. The nanobeams also support waveguide modes, which can be evanescently coupled with high efficiency to an optical fiber taper \cite{ref:michael2007oft}, as shown schematically in  Fig.\ \ref{fig:schematic}(b), by tuning the device geometry to match the nanobeam mode phase velocity with that of a fiber taper mode.  Despite the refractive index difference of the SiO$_2$ fiber taper and diamond, phase--matching is realized in nanobeams with subwavelength cross--section. This is illustrated in Fig.\ \ref{fig:schematic}(c)  which shows the effective refractive index dispersion, $n_\text{eff}(\lambda)$, of the fundamental TE modes of a fiber taper and of a diamond nanobeam, together with that of the ``supermodes'' of the evanescently coupled waveguides ($n^\pm_{\text{eff}}$) \cite{ref:yariv2006poe}. Phase--matching occurs at $\lambda_o \sim 1570\,\text{nm}$, where an anti--crossing in $n^{\pm}_\text{eff}(\lambda)$  indicates that the waveguide modes are coupled.

Evanescent waveguide coupling, which has been well studied in the context of photonic and optoelectronic devices \cite{ref:yariv2006poe, ref:barclay2003ede}, is shown here to  provide sensitive optomechanical readout of mechanical fluctuations which modulate the waveguide separation, $h$. In contrast to cavity-optomechanical systems which exploit  dispersive coupling between mechanical fluctuations and narrow-band optical resonances \cite{ref:aspelmeyer2013co},  waveguide optomechanical coupling utilizes wide-band dissipative optical transduction of the nanobeam mechanical position. The sensitivity of the evanescent waveguide optomechanical coupling can be derived from coupled-mode theory (Appendix \ref{sec:cm}), and is determined by the dependence of the normalized fiber taper transmission, $T$ on the separation $h$ between the waveguides:
\begin{equation}
T(h,\lambda)=\text{cos}^2(sL_c)+\left(\frac{\Delta\beta}{2}\right)^2  \frac{\text{sin}^2(sL_c)}{s^2}.
\label{eq:transmission}
\end{equation} 
 Here $L_c$ is the coupler interaction length determined by the fiber taper geometry,  $\Delta \beta(\lambda)=\Delta  n_{\text{eff}}(\lambda)2\pi/\lambda$ is the propagation constant mismatch of the uncoupled waveguide modes, $s^2 = \kappa^2 + \Delta\beta^2/4$. $\kappa(h)$ is the per--unit length amplitude coupling coefficient, which depends on the overlap of the evanescent fields of the coupled waveguides. The theoretical displacement sensitivity achievable by monitoring fluctuations in $T$ is
\begin{equation}\label{eq:Sx}
s_x^\text{o}(f) = \sqrt{\frac{2 T(h,\lambda) P_d \hbar\omega_o /\eta_{qe} + S_p^\text{d}}{P_d^2 \left(\partial T / \partial \kappa\right)^2\left(\partial \kappa / \partial h \right)^2\left(\partial h / \partial x \right)^2}},
\end{equation}
where $S_p^\text{d}$ is the single--sided noise equivalent optical power spectral density (units of $\text{W}^2/\text{Hz}$) of the detector and technical noise, and $P_d$ is the detected output power in absence of coupling ($T=1$).  The first term in the numerator accounts for photon shot noise, where $\eta_{qe}$ is the detector quantum efficiency.  This expression neglects radiation pressure backaction, which is not significant for the measurements presented here, but will ultimately limit $s_x^\text{o}$. The impact of coupler geometry, operating condition, and waveguide design on detection sensitivity is described by the denominator of Eq.\ \eqref{eq:Sx}. Sensitivity is maximized for mechanical resonances whose displacement $x$ efficiently modulates $h$, i.e., $|\partial h / \partial x| \sim 1$.   Strong evanescent overlap enhances $|\partial\kappa/\partial h|$, while phase--matching and operation near  $\kappa(h) L_c \sim \pi/4, 3\pi/4,...$ maximizes $|\partial T/\partial \kappa|$. 

Experimental observation of the optomechanical properties of the diamond waveguide optomechanical system was performed by measuring $T(t) =\overline{T} + \delta T(t)$ of a dimpled optical fiber taper \cite{ref:michael2007oft} positioned in the nanobeam near--field for varying $h$ and $\lambda$. The efficiency of the evanescent coupling determines the time--averaged transmission, $\overline{T}$. Fluctuations of the nanobeam position, together with other noise, are imprinted on $\delta T(t)$. Figure \ref{fig:waveguide_coupling}(a) shows $\overline{T}(\lambda)$ when the fiber taper is positioned at $h \sim 200\, \text{nm}$ above the center of a nanobeam ($w\times d = 460 \times 250\,\text{nm}^2$). Minimum transmission $\overline{T}_o = 0.05$, corresponding to coupling efficiency of $1 - \overline{T}_o = 95\%$ assuming negligible insertion loss, is observed near $\lambda_o = 1560\,\text{nm}$. The $3\,\text{dB}$ bandwidth $\Delta\lambda > 150\,\text{nm}$ is consistent with predictions from the $n_\text{eff}^\pm(\lambda)$  anti--crossing in Fig.\ \ref{fig:schematic}(c), and increasing $w$ was observed to increase $\lambda_o$ (Fig.\ \ref{fig:waveguide_coupling}(a) inset), consistent with expected behavior. 

The coherent nature of the waveguide coupling is revealed by  $\overline{T}_o(h)$. As shown in Fig.\ \ref{fig:waveguide_coupling}(b), $\overline{T}_o(h)$ is minimized at  $h \sim 200\,\text{nm}$, where $\kappa L_c =\pi/2$. For $h < 200$ nm, $\overline{T}_o$ increases with decreasing $h$ due to the co--directional coupling undergoing more than a half ``flop'' and light coupled into the nanobeam being out--coupled back into the fiber taper \cite{ref:barclay2003ede}.  In contrast, incoherent scattering loss increases monotonically with decreasing $h$ \cite{ref:basarir2012mtn}, and is small in the system studied here. As shown in Fig.\ \ref{fig:waveguide_coupling}(b), Eq.\ \eqref{eq:transmission}  fits $\overline{T}_o(h)$ well with an exponentially decaying $\kappa(h)$ as a fitting parameter, which in turn agrees closely with $\kappa(h)$ predicted from coupled--mode theory for interaction length $L_c \sim 7~\mu\text{m}$. This $L_c$ is consistent with the observed curvature in optical images of the fiber taper dimple.

The optomechanical properties of the coupled waveguides were observed from the power spectral density $S_v(f)$ of the photodetected signal generated by fluctuations in output power $P_d\delta T(t)$.  Figure \ref{fig:optomechanics}(a) shows the equivalent displacement spectral density $s_x(f)$ (units $\text{m}/\sqrt{\text{Hz}}$) of the nanobeam motion when the fiber taper is positioned $h \sim 100\,\text{nm}$ above the nanobeam.   Peaks from thermally driven nanobeam resonances are clearly visible at frequencies $f_m = \left[ 1.3,3.1,4.2 \right]\,\text{MHz}$, corresponding to the fundamental out--of--plane ($\mathbf{v}_1$), fundamental in--plane ($\mathbf{h}_1$), and second order out--of--plane ($\mathbf{v}_2$) resonances, whose simulated displacement profiles are shown in Fig.\ \ref{fig:schematic}(b). Resonance labels can be  determined by comparison of $f_m$ with simulations and by measuring transduction sensitivity as a function of in--plane taper position.  $s_x(f)$  was obtained by calibrating $S_v(f)$  to the theoretical thermomechanical power spectral density of the $\mathbf{v}_1$ resonance ($m = 7.6\,\text{pg}$), as described in Appendix \ref{sec:calib}.  The observed technical noise floor of  $s_x^o = 9.5\,\text{fm}/\text{Hz}^{1/2}$ is below that of other broadband integrated waveguide measurements \cite{ref:li2009bat,ref:basarir2012mtn, ref:rath2013doc}, and is more than an order of magnitude more sensitive than typical free--space reflection measurement techniques \cite{ref:burek2013nrs}.   This sensitivity can be further improved to the sub-$\text{fm}/\text{Hz}^{1/2}$ range by increasing $L_c$ and $P_d$.

The $h$ dependence of $S_v(f_m)^{1/2}$ and $s_x^o$ for the $\mathbf{v}_1$ resonance, shown in Figure \ref{fig:optomechanics}(b), provides a direct measure of the optomechanical coupling. The on--resonance signal is directly related to the slope of $\overline{T}(h)$: $S_v(h;f_m)^{1/2} \propto |d\overline{T}/dh|$, with a distinct minimum when $\kappa L_c =\pi/2$. Highest sensitivity is observed at $h = 100\,\text{nm}$ ($\kappa L_c = 3\pi/4$), and $s_x^o(h)$ agrees well with predictions from the experimentally characterized $\overline{T}(h)$ and Eq.\ \eqref{eq:Sx}.  The minimum $s_x^o(h)$ in Fig.\ \ref{fig:optomechanics}(b) is degraded compared to the sensitivity of the measurement in Fig.\ \ref{fig:optomechanics}(a) due to operating at lower $P_d$.   $d\overline{T}/dh$ is related to the effective dissipative optomechanical coupling coefficient of the system, defined by the change in waveguide tunneling rate of photons in the coupling region for a given change in $h$: $g_e/2\pi \sim c\,|\frac{dT_o}{dh}| /(2\pi n_{\text{g}} L_c\sqrt{T_o}) > 35\,\text{GHz/nm}$ where $c/n_g$ is the group velocity of light in the coupler.

%\begin{figure}
%\begin{center}
%\epsfig{figure=Figure3b_ver2.pdf, width=1.0\linewidth}
% \caption{ Measured $S_v(f)$ for a nanobeam with $L \times w\times d = 60 \times 0.75\times 0.3\,\mu\text{m}^3$ (a), in vacuum, (b), at 5K. Insets: fits to observed mechanical resonances, and corresponding $Q_m$.}
%\label{fig:high-Qm}
%\end{center}
%\end{figure}

Nanobeam mechanical dissipation in the measurements discussed above was dominated by damping from the ambient air environment. To assess the nanobeam mechanical properties, measurements were performed in vacuum and low--temperature conditions. Generally, after reducing air damping $Q_m$ was observed to increase above $10^4$ in all device geometries, with the lowest dissipation ($Q_m > 10^5$) observed in the larger cross--section ($w\times d=750\times300\,\text{nm}^2$) devices. Figures \ref{fig:optomechanics}(c) and \ref{fig:optomechanics}(d) show mechanical resonances of an ultrahigh--$Q_m$ nanobeam measured in vacuum (80 $\mu\text{Torr}$) and in cryogenic conditions (T = $5\,\text{K}$ and $5\, \mu\text{Torr}$).  In vacuum, the $\mathbf{v}_1$ and $\mathbf{h}_1$ resonances have $Q_m \approx 1.4\times10^5$ and $2.0\times 10^5$ respectively. At low temperatures, dissipation was further reduced, such that $Q_m \approx 1.7\times 10^5$ and $7.2 \times 10^5$ for the $\mathbf{v}_1$ and $\mathbf{h}_1$ modes, respectively. These values of $Q_m$ are higher than previous reports of nanomechanical devices fabricated from single crystal optical grade diamond \cite{ref:ovartchaiyapong2012hqf, ref:burek2013nrs, ref:tao2013scd}. All measurements were made at low power to avoid inducing optomechanical linewidth narrowing. Fabricating devices from electronic grade material or with larger dimensions may allow for further increased in $Q_m$, as in Ref.\ \cite{ref:tao2013scd}.  

Given the demonstrated device performance, the measurement precision required to reach the standard quantum limit is $s_x^\text{SQL} = \sqrt{\hbar Q_m/(m\omega_m^2)}=  3.1\,\text{fm}/\sqrt{\text{Hz}}$ where $m = 23\,\text{pg}$ is the effective mass of the $\mathbf{h}_1$ resonance \cite{ref:teufel2009nmm}.  While this is $\sim 3$ times smaller than the measurement sensitivity demonstrated here, promising approaches for reducing $s_x^o$ below $s_x^\text{SQL}$ include improving the optomechanical coupling by increasing $L_c$, and fabricating higher-$Q_m$ devices from electronic grade diamond material \cite{ref:tao2013scd}.

\section{Tunable nonlinear dynamics}

\begin{figure*}
\begin{center}
\epsfig{figure=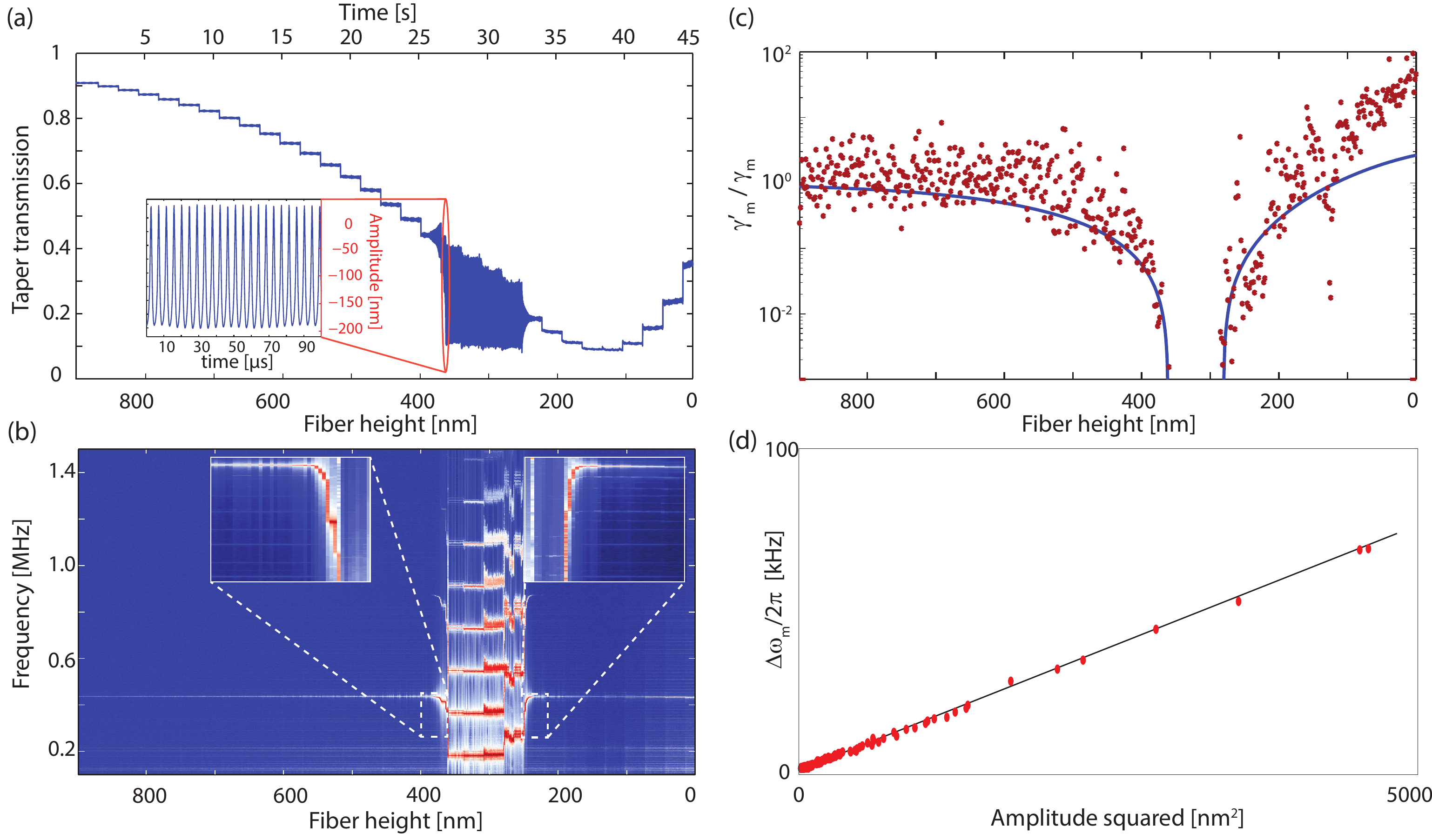, width=1\linewidth}
 \caption{Nanobeam self--oscillations. (a), Time resolved $T(h(t))$ of a phase-matched fiber taper--nanobeam ($L \times w\times d = 80 \times 0.50\times 0.25\,\mu\text{m}^3$) system as $h$ is reduced in discrete steps (visible as sharp steps in $T$). Inset: large amplitude oscillation of the nanobeam position.   (b), Spectrograph showing the power spectral density of the data in (a). (c), Predicted renormalized mechanical dissipation rate compared with measured values.  (d), Scatter plot of shift in nanobeam resonance frequency vs.\ oscillation amplitude squared for data in the inset regions from (b) at the onset of self--oscillation. The solid line is a linear guide to the eye.}
\label{fig:self}
\end{center}
\end{figure*}

Backaction from waveguide optomechanical coupling can dramatically modify the nanobeam dynamics, and is shown here to amplify the nanobeam motion and reveal nonlinear nanomechanical properties of the device.  These effects can be tuned by adjusting the waveguide position, and depend critically on the presence of internal stress in the nanobeam.  Demonstration of optomechanically modified nanobeam dynamics is shown in Fig.\ \ref{fig:self}(a), where time--resolved $T(t)$ of a fiber taper coupled to a high--aspect ratio nanobeam ($L\times w\times d = 80\times 0.48\times 0.25\,\mu\text{m}^3$) in a vacuum environment is recorded while $h$ is discretely stepped (30 nm/step, 0.8 steps/s, $P_{i} \sim 300\,\mu\text{W}$).  For large $h$, $T(h)$ behaves similarly to the ambient condition measurements in Fig.\ \ref{fig:optomechanics}(a).   However, at $h \sim 400\,\text{nm}$ fluctuations $\delta T(t)$ become large--amplitude self--oscillations with a peak-to-peak change $\Delta T > 0.3$, corresponding to nanobeam displacement exceeding 200 nm, as illustrated in the inset to Fig.\ \ref{fig:self}(a). To the best of our knowledge, this amplitude is larger than in previous reports of on-chip externally driven single crystal  \cite{ref:sohn2014das} and polycrystalline \cite{ref:rath2013doc} diamond nanobeams. Finite elements simulations predict a variation of axial stress at the center of the nanobeam on the order of $70\,\text{MPa}$ for the $\mathbf{v}_1$ resonance of the compressed nanobeam oscillating with this amplitude.  When $h < 225\,\text{nm}$, the self--oscillations stop.   Figure \ref{fig:self}(b)  shows a spectrograph of this data, where the $\mathbf{v}_1$ resonance near $f_m = 430\,\text{kHz}$ with $Q_m = 2.5\times 10^4$ is observed to increase in amplitude, shift to lower frequency, and generate harmonics.  

In contrast to diamond nanobeam oscillations driven by external actuation \cite{ref:sohn2014das} or resonant optical modulation \cite{ref:rath2013doc}, the observed self--oscillations are driven by the dynamic interaction between a delayed photothermal force \cite{ref:metzger2004ccm, ref:favero2007ocm} and the waveguide optomechanical coupling.  The resulting optomechanical backaction renormalizes the mechanical dissipation rate from $\gamma_m$  to $\gamma_m'$,
\begin{equation}\label{eq:gamma_mp}
\frac{\gamma_m'}{\gamma_m } =   1- Q_m\frac{\mathcal{F}}{k}\frac{\omega_m\tau}{1+\omega_m^2\tau^2}\, \frac{dT}{dx} P_i \zeta L_i,
\end{equation}
where $\tau$ is photothermal force response time, and $\mathcal{F}$ is the photothermal force strength per unit absorbed power $P_{\text{abs}}=(1 - T)P_i\zeta L_i$, defined by the corresponding nanobeam deflection and spring constant  $k = m \omega_m^2$ of the mode of interest.  $L_i$ is the distance over which input light propagates in the nanobeam before being out--coupled, and $\zeta$ is the waveguide absorption coefficient per unit length.  Figure \ref{fig:self}(c) compares experimentally observed $\gamma_m'(h)$ with theoretical values obtained from Eq.\ \eqref{eq:gamma_mp} input with measured $T(h)$ as well as parameters discussed below. Predicted $\gamma_m'(h)$ agrees well with experiment, particularly in reproducing the range of $h$ over which the optomechanical coupling ($|dT/dx|$) is sufficiently large such for $\gamma_m' < 0$, resulting in self--oscillations. Note that disagreement for $h < 100\,\text{nm}$ is in part related coupling between fluctuating fiber position and $\omega_m$, which becomes significant compared to $\gamma_m$ for small $h$ (see Appendix \ref{sec:photothermal}.)

%For $h < 200\,\text{nm}$,  photothermal damping broadens $\gamma_m$. However, coupling between $\omega_m$ and the fiber taper position for this range of $h$, combined with fluctuations in fiber position due to external vibrations, artificially broadens $\gamma_m$ and makes accurate measurement the cooling rate challenging.

In addition to exhibiting ultrahigh--$Q_m$  and strong dissipative optomechanical coupling, the nanobeams have two properties which  make them sensitive to optomechanical photothermal actuation. The nanobeam thermal time constant $\tau\sim 0.7\,\mu\text{s}$, calculated using finite element simulations, is on the same timescale as $\omega_m$. As seen from Eq.\ \eqref{eq:gamma_mp}, this is a necessary condition for significant feedback from the photothermal force \cite{ref:metzger2004ccm, ref:favero2007ocm, ref:barton2012pso, ref:zaitsev2012ndm}.  A more subtle but equally important property is the presence of compressive stress and accompanying buckling in the nanobeam, which as discussed below, dramatically enhances $\mathcal{F}$.   

For the nanobeam geometry studied here, finite element analysis presented in Appendix \ref{sec:photothermal} indicates that $\mathcal{F}$ can be enhanced by over two orders of magnitude in a compressed and buckled nanobeam.  The level of compressive stress in a nanobeam can be estimated from the deviation of $\omega_m$ from the nominal value expected for an uncompressed device \cite{ref:burek2013nrs}. The nanobeam studied in Fig.\ \ref{fig:self} is observed to be in a post-buckled state with measured $\omega_m/2\pi \sim 430\,\text{Hz}$ significantly lower than the nominal value expected ($680\,\text{kHz}$) for an uncompressed nanobeam. Matching finite element simulated and observed $\omega_m$ predicts an internal compressive stress of $\sim 37\,\text{MPa}$ and an accompanying buckling amplitude of $\overline{x} = -122\,\text{nm}$.  For this value of internal stress and buckling, a photothermal force of $\mathcal{F} = -26\,\text{pN}/\mu\text{W}$ is predicted, which is over 100 times larger than predicted for an uncompressed device.   Note that the negative signs of $\mathcal{F}$ and $\overline{x}$ indicate that the photothermal force and buckling amplitude are in the down direction, consistent with all of the self-oscillation behavior discussed in this section.

Given the above device parameters, the only free fitting parameter needed to match Eq.\ \eqref{eq:gamma_mp} with experiment is the waveguide absorption coefficient, which is found to be $\zeta = 0.12\,\text{cm}^{-1}$, corresponding to an optical loss rate effective quality factor of $Q_o \sim 6.6\times 10^5$. This absorption rate is consistent with loss in other diamond nanophotonic devices \cite{ref:burek2014hqf}, and indicates that only a fraction ($\sim 100 \,\text{nW}$) of the input power is absorbed and responsible for driving the self--oscillations.

Nonlinear coupling between nanobeam oscillation amplitude and $\omega_m$ provides an additional probe of the internal stress and buckled state of the device. Softening of the mechanical resonance frequency by $\Delta\omega_m(h)$ is observed at the onset of self--oscillations, as highlighted in  Fig.\ \ref{fig:self}(b). Nanobeam softening and hardening is a well known indicator of internal stress \cite{ref:sohn2014das, ref:rath2013doc}, and the softening observed here is found to be closely related to the buckled nanobeam state. Nanobeam buckling breaks the device vertical symmetry and introduces a nonlinear softening term to the nanobeam dynamics which counteracts the hardening effect described by the intrinsic Duffing nonlinearity \cite{ref:kozinsky2006tnd}. This competition between nonlinear effects is given by,
\begin{equation}\label{eq:duffing}
\Delta \omega_m = \frac{v^2}{\omega_m}(\frac{3}{8}\alpha_3 - \overline{x}^2	\frac{15}{4\omega_m^2}\alpha_3^2),
\end{equation} 
where $v$ is the oscillation amplitude, and $\alpha_3$ is the Duffing coefficient of the unbuckled nanobeam as described in Appendix \ref{sec:photothermal}. Equation \eqref{eq:duffing} clearly shows the softening influence of $\overline{x}$. The optomechanical system studied here provides a direct measurement of $\Delta\omega_m$ and $v$ for varying $h$, allowing $\overline{x}$ to be estimated experimentally. Figure \ref{fig:self}(d) shows a scatter plot of measured $\Delta\omega_m$ as a function of $v^2$ for varying fiber taper position. As predicted from Eq.\ \eqref{eq:duffing}, $\Delta\omega_m$ and $v^2$ are found to be linearly related, and from this data and Eq.\ \eqref{eq:duffing}, $\overline{x} = -98\,\text{nm}$ is estimated, which is in excellent agreement with finite element simulation predictions of $\overline{x}$ given above. 

During the self--oscillation limit--cycle, a low fundamental self--oscillation frequency of $\sim 180\,\text{kHz}$ is measured, consistent with the behavior of a nanomechanical resonator oscillating between buckled states, as observed by Bagheri et al.\ \cite{ref:bagheri2011dmn}.  Further evidence of this behavior is found in Fig.\ \ref{fig:self}(a), which shows that during self--oscillations $\overline{T}$ decreases, indicating that the nanobeam, initially in a buckled down state, moves on-average closer to the fiber taper while self--oscillating. Note that harmonics which emerge during self--oscillation, becoming stronger and more numerous as $v$ increases, result from both the nonlinear response of $\overline{T}(h)$ and nonlinearities of the nanobeam \cite{ref:poot2012bls}. 

\section{Discussion and conclusion}

The  waveguide optomechanical interface demonstrated here possess a unique combination of high sensitivity, broad bandwidth, high quality single-crystal diamond material, and high--$Q_m$, and has potential to allow measurement of quantum motion of nanobeam resonances. Excitation of nanomechanical self--oscillations with nW absorbed power illustrates the sensitivity of the diamond nanobeams to small driving forces, and their nonlinear dynamical softening provides a glimpse of the changing stress within the nanobeam during large amplitude oscillations. This demonstration of optomechanical excitation of the diamond nanomechanical environment is a step towards optomechanical control of quantum electronic systems such as nitrogen vacancies \cite{ref:arcizet2011snv, ref:bennett2013pis,ref:macquarrie2013msc, ref:ovartchaiyapong2014dsc, ref:teissier2014scn}. It is promising for implementations of nanomechanical logic in diamond \cite{ref:bagheri2011dmn}, and can be extended to use the optical gradient force for optomechanical nanobeam actuation, enabling excitation of  higher frequency resonances of smaller structures  \cite{ref:eichenfield2009oc}.   Finally, the scalable nanofabrication technique demonstrated here is widely applicable to diamond nanophotonic devices for sensing, nonlinear optics, and quantum information processing, and can be easily adopted by researchers with access to standard nanofabrication tools. Extending this approach to integrate an optical cavity into the waveguide optomechanical system, for example by patterning the waveguide ends with mirrors \cite{ref:burek2014hqf}, will allow studies of dissipative optomechanical cooling  of the nanobeam resonances in the unresolved--sideband regime \cite{ref:elste2009qni}.

\subsection*{Acknowledgements}
We thank Charles Santori and David Fattal for useful initial discussions related to the fabrication approach used here. We would like to acknowledge support for this work from NSERC, iCore/AITF, CFI, NRC.

\appendix

\section{Fabrication process}\label{sec:fab}

The fabrication process flow is outlined in Figure 2.  A chemical-vapor-deposition-grown, $\langle 100 \rangle$-oriented single crystal diamond optical--grade substrate (Element 6) is cleaned in boiling piranha (3:1 H$_2$SO$_4$:H$_2$O$_2$) and coated with a 300-nm-thick layer of PECVD Si$_3$N$_4$ as a hard mask.  Next, $2-5\,\text{nm}$ of titanium is deposited as an anti-charging layer.  Nanobeam structures with axes aligned along the $\langle 110 \rangle$ direction are patterned in ZEP520A electron-beam lithography resist and developed at $-15$ $^\text{o}$C in ZED-N50.   The resulting ZEP pattern is transfered to the nitride hard mask using an inductively-coupled plasma reactive-ion etch (ICPRIE) process with C$_4$F$_8$/SF$_6$ chemistry.  An anisotropic oxygen plasma ICPRIE step transfers the pattern to the diamond, followed by a 160-nm-thick conformal coating of PECVD Si$_3$N$_4$ to protect the vertical diamond sidewalls.  A short anisotropic C$_4$F$_8$/SF$_6$ ICPRIE step clears the nitride from the bottom of the windows while keeping the side and top surface of the devices protected.  To create suspended nanobeams, a quasi--isotropic oxygen plasma etch \cite{ref:hwang2004nep} is performed at 2500 W ICP power, 0 W RF, and elevated wafer temperature of 250 $^\text{o}$C.  This etch step is relatively slow, requiring $\sim 5$ hours to undercut the nanobeams studied here.  The undercut etch rate was observed to increase at higher--ICP power, however this option was not available for the devices fabricated for this report.  An Oxford Plasmalab  is used for all plasma etch steps.  Finally, the titanium and nitride layers are removed by wet--etching in 49\% HF, and the sample is cleaned a second time in boiling piranha.   Note that adding an second vertical diamond etching step immediately prior to the quasi--isotropic etch is expected to reduce the necessary undercut time, as in the related SCREAM silicon process \cite{ref:shaw1994scr}.

\section{Evanescent coupling to diamond nanobeams}\label{sec:cm}
Numerical prediction of the coupling coefficient $\kappa(h)$ describing the interaction between the nanobeam and fiber taper waveguide modes can be obtained from the optical dispersion of the eigenmodes of the coupled waveguides  (``supermodes''). This process is described below.

\begin{figure}
\begin{center}
\epsfig{figure=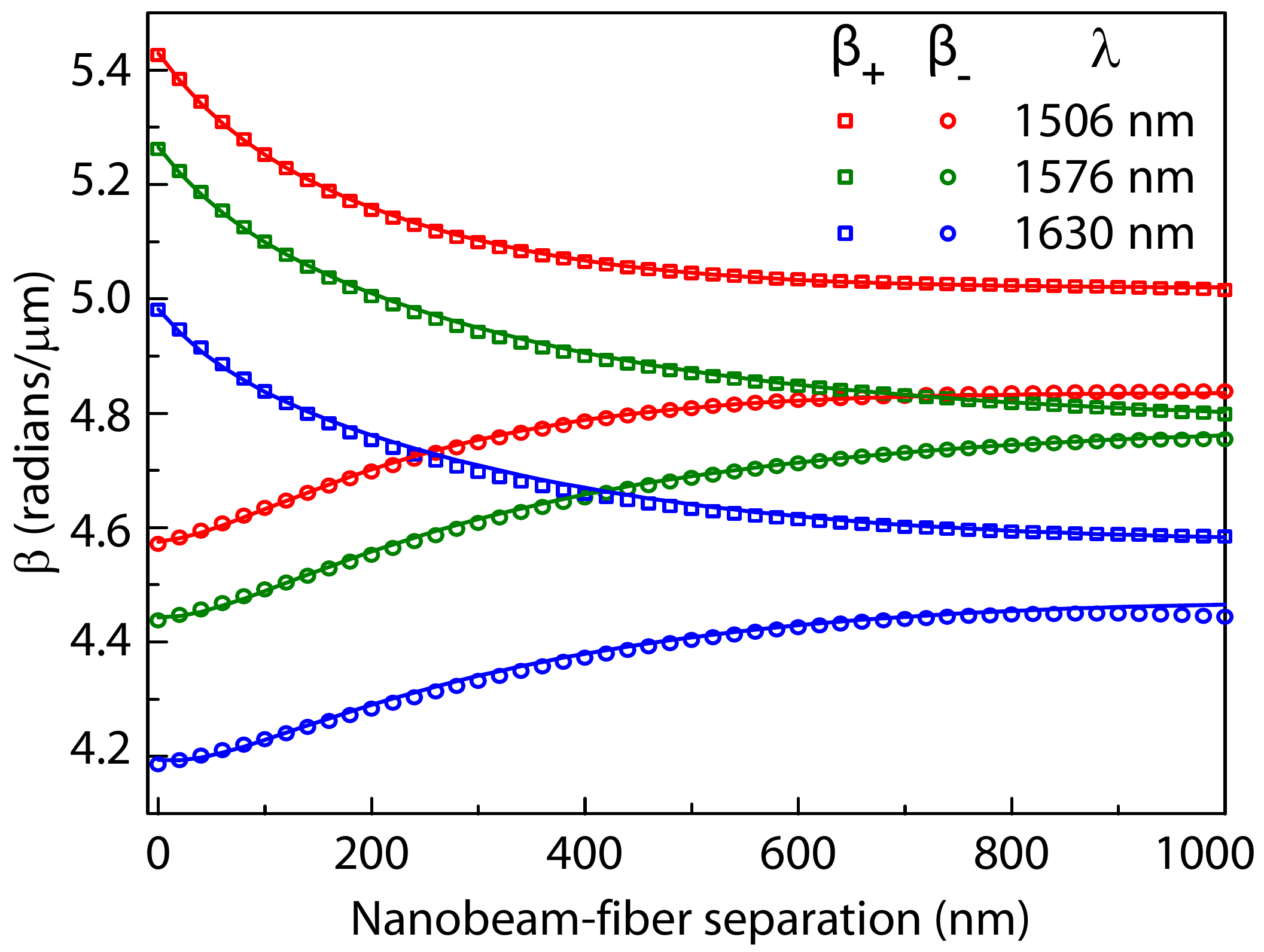, width=1\linewidth}
 \caption{Propagation constants ($\beta_{\pm}$) of the even and odd supermodes of the coupled nanobeam and fiber taper waveguides, as a function of waveguide separation $h$, for varying $\lambda$.  Points indicated by open squares  and circles where calculated using Lumerical MODE Solutions. Solid lines are fits using the coupled mode theory model described in this section.}
\label{fig:beta_gap-sweep}
\end{center}
\end{figure}

The field propagating through the coupled waveguide system can be represented as a superposition of modes of the uncoupled optical fiber taper and diamond nanobeam. In the case of waveguides with two nearly phase--matched co--propagating (positive group velocity) modes, this evolution can be approximately described by coupled mode equations describing the field amplitude $a_\text{f,n}(z)$ in the fiber and  nanobeam waveguides, respectively, as a function of propagation distance $z$ through the coupling region:
\begin{align}
\frac{da_\text{f}}{dz} &= -j(\beta_\text{f} +\kappa_\text{ff}) a_\text{f} - j\kappa a_\text{n}, \label{eq:da}\\
\frac{da_\text{n}}{dz} &= -j(\beta_\text{n} + \kappa_\text{nn}) a_\text{n} - j\kappa a_\text{f}.\label{eq:db},
\end{align}
where $\beta_\text{f}$ and $\beta_n$ are the $\lambda$ dependent propagation constants of the uncoupled fiber and nanobeam modes, respectively.  Also included in this model are ``self-term'' corrections $\kappa_\text{ff,nn}(h)$ to $\beta_\text{f,n}$ resulting from the modification of the local dielectric environment by the coupled waveguides. From Eqs.\ \eqref{eq:da} and \eqref{eq:db},  supermodes of the coupled waveguides, can be found with $z$ dependence $e^{-i\beta_\pm z}$, where
\begin{align}
\beta_{\pm}=\frac{\tilde{\beta}_\text{f}+\tilde\beta_\text{n}}{2}\pm\sqrt{\left(\frac{\tilde\beta_\text{f}-\tilde\beta_\text{n}}{2}\right)^2+\kappa(h)^2},
\label{eq:beta1}
\end{align}
with $\tilde\beta_\text{f,n} = \beta_\text{f,n}(\lambda)+\kappa_\text{ff,nn}(h)$ \cite{ref:yariv2006poe}.  Supplementary Figure \ref{fig:beta_gap-sweep} shows $\beta_{\pm}(h,\lambda)$ calculated with a mode solver  (Lumerical MODE Solutions).   $\kappa(h)$ and $\kappa_\text{ff,nn}(h)$ were then determined by fitting $\beta_{\pm}(h,\lambda)$ with  Eq.\ \eqref{eq:beta1}, where numerically calculated  $\beta_\text{f,n}(\lambda)$ of the uncoupled waveguides are known input parameters.  The resulting values for $\kappa(h)$ and $\kappa_\text{ff,nn}(h)$, shown in Supplementary Fig.\ \ref{fig:kappa_gap-sweep}, are found to decay exponentially with $h$. For simplicity, it was assumed that $\kappa_\text{ff} =\kappa_\text{nn}$ during the fitting process.   Note that while $\kappa(h)$ decays over a length--scale comparable to the waveguide evanescent field decay, $\kappa_\text{ff,nn}(h)$ are near--field terms which decay much more quickly, and do not significantly impact the predicted values of $\kappa(h)$ for $h > 100\,\text{nm}$.  For a given $h$, $\beta_{\pm}(\lambda)$ can be converted to an effective index dispersion curve, $n_{\pm}(\lambda) = \beta_{\pm}(\lambda)\lambda/2\pi$, as shown in  Fig.\ 1(c), which clearly exhibits an anti--crossing near $\beta_\text{f} = \beta_\text{n}$ whose width scales with $|\kappa|$.  The excellent agreement between the numerically calculated $\beta_{\pm}$ and the semi-analytic model described by Eq.\ \eqref{eq:beta1} indicates that this model is suitable for predicting the waveguide coupling. 

The coupler response can also be predicted from this coupled mode analysis. Solving Eqs.\ \eqref{eq:da} and \eqref{eq:db} with boundary condition corresponding to unity input power to the fiber taper ($|a_\text{f}(0)|^2 = 1, |a_\text{n}(0)|^2 = 0$), results in the taper transmission $T(\lambda,h)$ given by Eq.\ (1) in the main text. 

\begin{figure}
\begin{center}
\epsfig{figure=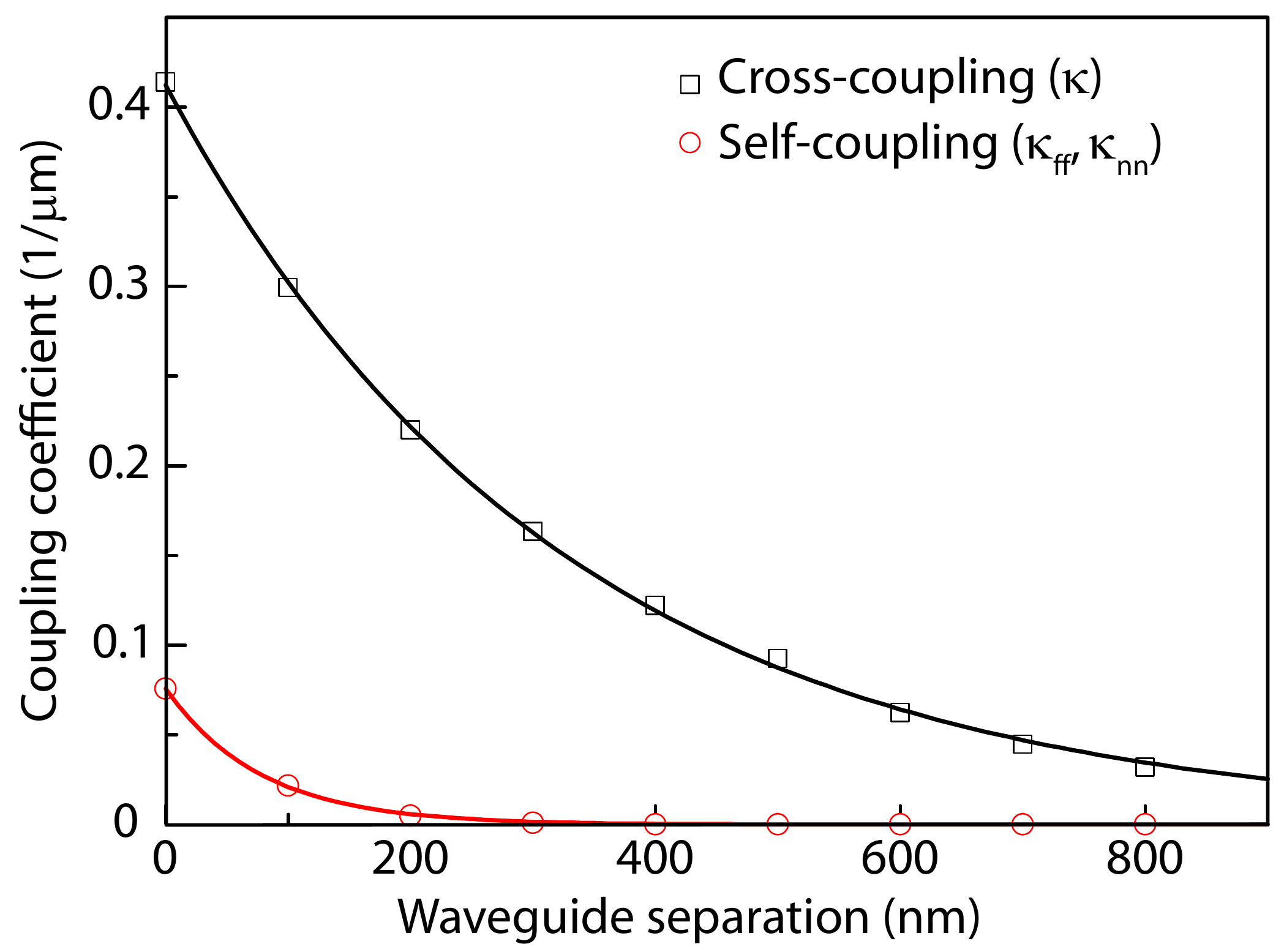, width=1\linewidth}
 \caption{Waveguide coupling coefficients as a function fiber taper and nanobeam waveguide separation ($h$). Each data point was obtained by fitting $\beta_\pm(\lambda)$ with the coupled mode theory model. Solid lines are single exponential fits to the data points.}
\label{fig:kappa_gap-sweep}
\end{center}
\end{figure}
\section{Measurement setup}

The optomechanical properties of the diamond nanobeam waveguides were studied by monitoring the optical transmission of a dimpled optical fiber taper positioned in the near field of devices of interest. The dimpled fiber taper was fabricated by modifying the procedure presented in Ref.\ \cite{ref:michael2007oft}  to use a cermanic mold for creating a dimple.  Measurements performed in ambient conditions used high--resolution (50 nm) DC stepper motors to position the fiber taper.  Vacuum (room temperature) and cryogenic measurements were performed in a closed cycle cryostat (Montana Systems Nanoscale Workstation) whose sample chamber is configured with stick--slip and piezo positioning stages (Attocube) for controlling the sample and fiber taper positions. Before cooling, a turbo pump was used to evacuate the chamber to pressures in the $10^{-5}$ Torr range. Room--temperature vacuum measurements were performed in these conditions.  At 5K, cryo--pumping reduced the chamber pressure to the $10^{-6}$  Torr range.  During the low temperature measurements, the fiber taper was positioned in contact with lithographically defined supports on the diamond chip to reduce coupling of vibration from the cryostat cooling stages to low frequency resonances of the optical fiber taper. These supports, visible in Supplementary Fig.\ 4, allow the fiber taper to be positioned in the nanobeam near--field without contacting the nanobeam.

Two external cavity tunable diode lasers (New Focus Velocity 6700) were used to probe the fiber taper transmission $T$ over a wavelength range from 1475 -- 1625 nm.  A New Focus 1811 photodetector (PD1) with a noise equivalent power $S_P^{(\text{det})} = 2.5~\text{pW}/\sqrt{\text{Hz}}$ was used to monitor the average ($\overline{P}_d$) and fluctuating ($\delta P_d(t)$) output power from the fiber taper. A New Focus 1623 detector (PD2) was also used in some measurements of $\overline{P}_d$.

 A Tektronix RSA5106A real time spectrum analyzer (RSA) allowed fast spectral analysis during the experiments, and recording of IQ  time--series of the PD1 output voltage $V(t)$.  All of the $S_v(f)$ data presented here was generated in post-processing from $V(t)$ data.  By choosing a low center (demodulation) frequency (typically 0 or 2.5 MHz), and sampling $V(t)$ with a bandwidth exceeding the nanobeam resonance frequencies (typically 5 MHz) both $\delta T(t)$ and $\overline{T}$ could be recorded by the RSA.  To avoid damaging the RSA with a large DC input, a pair of bias-Ts (Minicircuits ZFBT-6GWB+) together with electrical attenuators were used to reduce the low frequency ($< 100$ kHz) components of $V(t)$.  The self--oscillation data in  Fig.\ 5 was then acquired in a single 45 s time--series while the fiber height above the sample was stepped in 30 nm increments. Attenuation of the low--frequency signal was compensated for in post--processing.

\section{Optomechanical transduction sensitivity}\label{sec:calib}
This section describes the procedure for predicting the theoretical measurement sensitivity of the optomechanical waveguide readout, and for extracting the experimentally observed measurement sensitivity from measured thermomechanical signals.  Single--sided power spectral densities are used throughout.
\subsubsection{Theoretical sensitivity}
The displacement sensitivity of the waveguide--optomechanical system is determined by the minimum mechanical motion to actuate a signal larger than the noise floor of the measurement apparatus. For direct photodetection of the optical power transmitted by the waveguide, the power spectral density, $S^{(s)}_{v}(f)$, of the transduced signal from a mechanical displacement described by spectral density $S_x(f)$ is
\begin{equation}\label{eq:G}
S^{(\text{s})}_v(f) = S_x(f)\left(g_\text{ti} P_d \frac{\partial T}{\partial x}\right)^2,
\end{equation}
where  $g_\text{ti}$ is the detector transimpedance gain, $P_d$ is the detected power in absence of coupling ($T = 1$), and $|\partial T/\partial x|$ describes the optomechanical actuation of the coupler. Similarly, the measurement noise can be written as
\begin{equation}
S^{(\text{n})}_v(f) = S^{(\text{SN})}_v(f) + S^{(\text{det})}_v(f)
\end{equation}
$S^{(\text{det})}_v$ describes noise intrinsic to the detector, and is related to the detector's noise equivalent power figure ($S^{(\text{det})}_P$) by $S^{(\text{det})}_v = g_\text{ti}^2 S^{(\text{det})}_P$. The contribution from photon shot noise is given by
\begin{equation}
S^{(\text{SN})}_v(f) = g_\text{ti}^2 \frac{2\hbar\omega_o T P_d}{\eta_{qe}}.
\end{equation}
where $\eta_{qe}$ is the detector quantum efficiency. Note that the impact of shot noise is affected by the operating point, $T$, of the coupler. This analysis does not consider optomechanical back--action, which is small for the optical powers used in the high sensitivity measurements presented here.   

To calculate the minimum detectable spectral density, $S_x^o$, we require unity signal to noise ratio, $S^{(\text{s})}_v = S^{(\text{n})}_v$, resulting in 
\begin{equation}\label{eq:sens}
S^{o}_x = \frac{2 \hbar\omega_o T P_d/\eta_{qe} +  S^{(\text{det})}_P}{\left( P_d \frac{\partial T}{\partial x}\right)^2}.
\end{equation}
$S^{o}_x$ has units of $\text{nm}^2/\text{Hz}$ and is related to the minimum  detection sensitivity given in the text by $s_x^o = \sqrt{S_x^o}$.
In the case of a co--directional evanescent coupler, $\partial T/\partial x$ can be calculated from Eq.\ (1) in the main text.  For perfect phase--matching ($\Delta\beta = 0$), 
\begin{equation}
\frac{\partial T}{\partial x} = - \sin(2\kappa L) \frac{\partial \kappa}{\partial h}\frac{\partial h}{\partial x}.
\end{equation}
Alternately, $\partial T/\partial x$ can be measured experimentally by recording $T(h)$ and determining $dT/dh$. This approach was used together with Eq.\ \eqref{eq:sens} to generate the predicted measurement sensitivity in Fig.\ 4, assuming that $\partial h/\partial x = -1$.

\subsubsection{Thermomechanical calibration}
The observed thermal nanobeam motion was used to calibrate the measurement noise floor using a standard procedure described in, for example, Refs.\ \cite{ref:hauer2013gpt,ref:wu2014ddo}.  Thermomechanical resonances in $S_v(\omega)$ were fit using $S_v(\omega) = S_v^{(n)}  + G S_x^{th}(\omega)$ where $S_x^{th}(\omega)$ is the single--sided power spectral density of a thermal oscillator,
\begin{equation}
S_x^{th}(\omega) = \frac{4k_B T_e\omega_m}{Q_m}\frac{1}{m\left[(\omega-\omega_m^2)^2+(\omega\omega_m/Q_m)^2\right]},
\end{equation}
and $G$ is a constant determined by the transduction gain of the measurement, as described theoretically by Eq.\ \eqref{eq:G}, and treated as a fitting parameter for the purpose of the calibration procedure.  Here $k_B$ is Boltzmann's constant, $T_e$ is the operating temperature, and $m$ is the effective mass of the resonance, as defined in Ref.\ \cite{ref:eichenfield2009apn}.  From the fit values, spectra can be converted from electrical ($\text{W}/\text{Hz}$) to displacement ($\text{m}^2/\text{Hz}$) power spectral density: $S_x = S_v/G$.

\section{Optical gradient force}

\begin{figure}
\begin{center}
\epsfig{figure=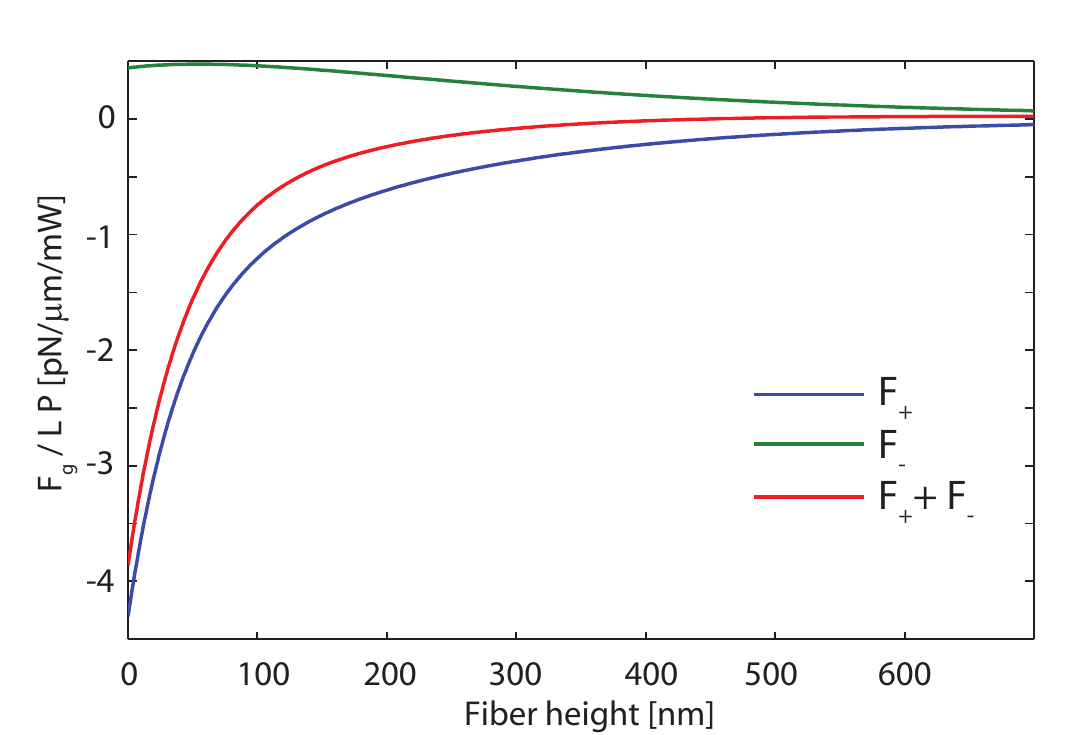, width=1\linewidth}
 \caption{Optical gradient force as a function of $h$, when the fiber taper is aligned along the nanobeam axis, and offset $500\,\text{nm}$ laterally, approximating the position in the self--oscillation measurements.}
\label{fig:fg}
\end{center}
\end{figure}

Dielectric objects in an evanescent field experience an optical gradient force.  The optical gradient force between coupled waveguides can be predicted from $n_{\pm}(h)$ using the formalism of Povinelli et al.\ \cite{ref:ma2011mkn}.  Following this formalism, the force induced by power $P_\pm$ in each of the supermodes is given by
\begin{equation}
F_{\pm} = \frac{P_\pm L_c}{c}\left.\frac{\partial n_{\pm}}{\partial h}\right\vert_{\omega_o} = -\frac{P_\pm L_c}{c}\left.\frac{\partial n_{\pm}}{\partial x}\right\vert_{\omega_o}.
\end{equation}
Operating at phase matching with power $P_i$ input into the fiber taper, the power in the even and odd supermodes is $P_\pm = P_i/2$, and the total optical gradient force on the nanobeam is 
\begin{equation}
F_{g} = \frac{P_i L_c}{2c}\left(\frac{\partial n_{+}}{\partial h} + \frac{\partial n_{-}}{\partial h}\right)
\end{equation}
where positive (negative) $F_{g}$ indicates a repulsive (attractive) force.  Supplementary Figure \ref{fig:fg} shows the predicted $F_{g}(h)/L_c\,P_i$, for varying alignment of the fiber taper with the center axis of the nanobeam.  Corrections due to the curvature of the dimpled fiber taper are not considered explicitly, and are assumed to be accounted for by the effective coupler length $L_c$ extracted from the experimental measurements.  Given this approximation, $F_g$ is constant over the  interaction length of the ideal two--port  waveguide coupler, and vanishes outside of the coupling region.

\section{Nanobeam photothermal and nonlinear dynamics}\label{sec:photothermal}
Waveguide-optomechanical coupling  was observed to induce nanobeam self--oscillations, as shown in Fig.\ 5 of the main text. Here we analyze this effect by modeling the nanobeam as a nonlinear harmonic oscillator interacting with the optical field of the coupled waveguides through a dynamic photothermal force and an instantaneous optical gradient force \cite{ref:metzger2008osc,ref:li2008hof, ref:restrepo2011cqt, ref:zaitsev2012ndm}. 

\subsubsection{Nanobeam equation of motion}

The dynamics of the nanobeam resonance driven by a stochastic thermal force, $F_s(t)$, and coupled to the optical fiber taper, are approximately described by the following equation of motion:
\begin{multline}\label{eq:eom}
\ddot{x} + \frac{\omega_m}{Q_m}\dot{x} + (\omega_m^2(\overline{x}) + \alpha_3 x^2 + \alpha_2 x) x  = \\
\frac{1}{m}\left( F_{g}(x) + F_{pt}(x,t) +  F_{s}(t)\right),
\end{multline}
where $x(t)$ is the displacement amplitude of the nanobeam mechanical resonance of interest, defined relative to the position of the undriven nanobeam. Changes in $x$ modulate the nanobeam and the fiber taper spacing $h$, and for the case of the $\mathbf{v}_1$ resonance, $h(x) = h_o - x(t)$ if the fiber taper is positioned $h_o$ above the center of the nanobeam. Optomechanical coupling arises from the dependence of the optical gradient force, $F_{g}(h(x))$, and the photothermal force,  $F_{pt}(h(x))$, on nanobeam position. In addition, $\omega_m$ varies due to internal strain resulting from optically induced changes in static deflection, $\overline{x}$, and local nanobeam temperature relative to the environment, $\Theta$. For sufficiently large $|x(t)|$, the nanobeam response becomes nonlinear, as described by $\alpha_2$ and $\alpha_3$.

\subsubsection{Thermal effects}
Local heating of the nanobeam by waveguide optical absorption occurs on a timescale determined by the nanobeam geometry and material properties. The thermal dynamics of the nanobeam, defined by the maximum temperature change, $\Theta$, relative to the operating temperature, are assumed to follow
\begin{equation}\label{eq:temp}
\frac{d\Theta}{dt} = -\kappa_{t}\Theta+ R L_i\zeta P_n,
\end{equation}
where $\tau = 1/\kappa_{t}$ is the nanobeam thermal time-constant and $P_n$ is the optical power coupled into the nanobeam. In absence of delayed optical feedback (i.e., a cavity), $P_n$ instantaneously follows $x$, and is given by
\begin{equation}
P_n = (1 - T(h(x)))P_i,
\end{equation}
with $T(h)$ described by Eq.\ (1) in the main text.  Here $\zeta$ is the per--unit length absorption coefficient of the nanobeam waveguide, and $R$ is the heating power per unit of absorbed optical power. Ideally, $\zeta$ is determined by the material properties of diamond, but in nanophotonic devices it can be modified by imperfect surfaces. The effective nanobeam waveguide optical interaction length is given by $L_i(h) = \int_0^L |a_n(z)|^2 dz/|a_f(0)|^2$, and represents the distance over which light propagates in the nanobeam.  It can be calculated from Eqs.\ \eqref{eq:da} and   \eqref{eq:db}.

Local heating of nanobeam waveguides induces deflections in the nanobeam position. We write the corresponding photothermal force as
\begin{equation}
F_{pt}(x,t) = \mathcal{F} \Theta(x,t) \frac{\kappa_{t}}{R},
\end{equation}
where $\mathcal{F}$ is the force per unit absorbed power ($L_i\zeta P_n$) in steady state. The resulting change in nanobeam deflection is $\delta\overline{x}_{pt} = F_{pt}/k$ where $k = \omega_m^2 m$ is the nanobeam spring constant. 

Local heating and accompanying thermal expansion of the device also modifies $\omega_m$. This effect is represented by
\begin{equation}
\frac{d\omega_{m}}{d\Theta}= C_t\frac{\kappa_t}{R}
\end{equation}
where $C_{t}$ is a constant related to the elastic properties of the nanobeam, and has units $\text{rad}\text{s}^{-1}\text{W}^{-1}$.

\begin{figure*}
\begin{center}
\epsfig{figure=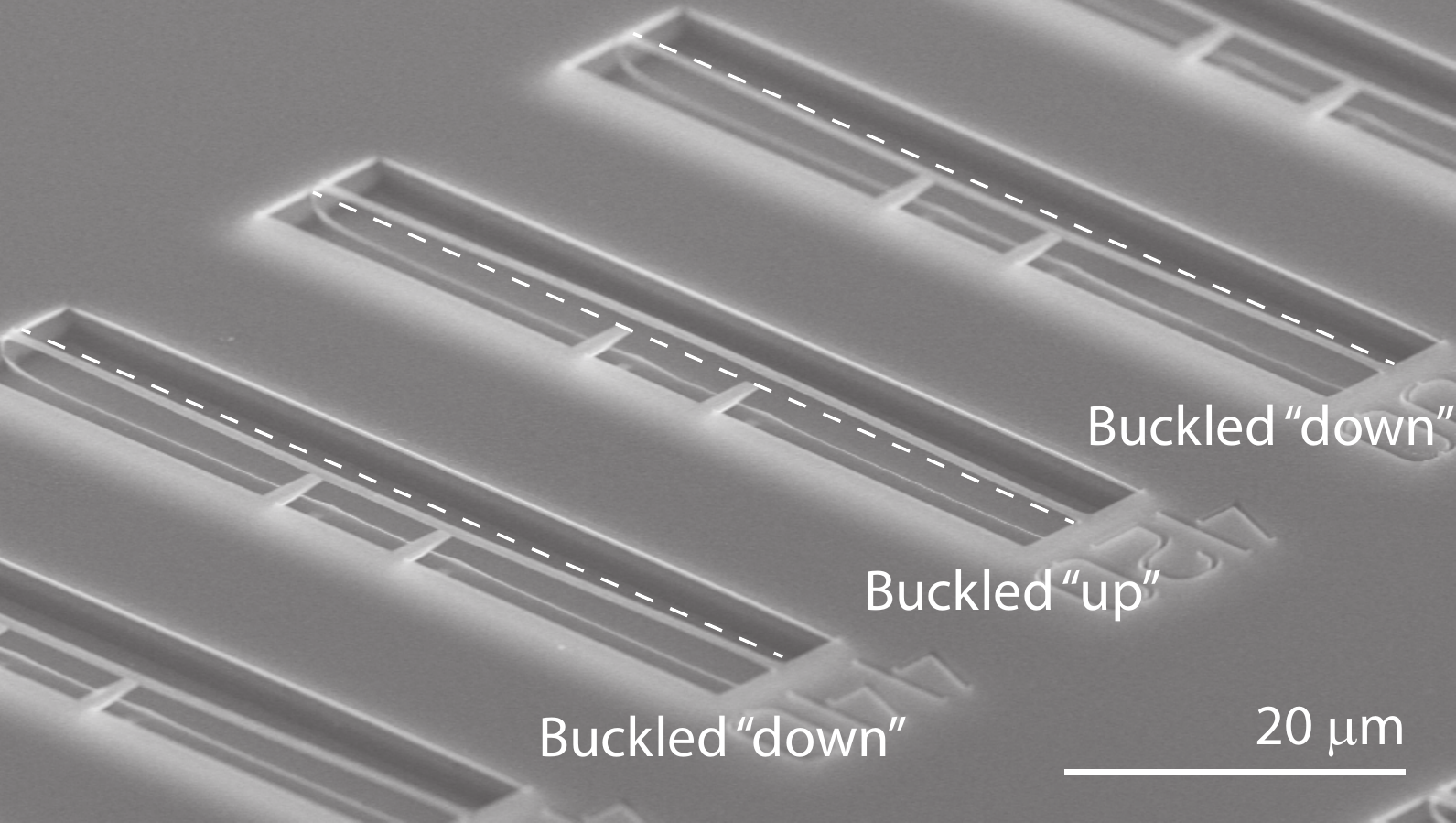, width=1\linewidth}
 \caption{SEM image of typical nanobeams observed to self--oscillate. Dashed straight lines are guides to emphasize the direction of nanobeam buckling.}
\label{fig:buckling}
\end{center}
\end{figure*}

\subsubsection{Static response of compressed nanobeams}

In general, $C_t$ and $\mathcal{F}$ sensitively depend on both the nanobeam elastic properties and geometry, and the internal residual stress acting on the device.  The nanobeams used in the self--oscillation studies have significant internal stress, manifesting in smaller $\omega_m$ than expected from their nominal dimensions, and can be in a buckled geometry as shown in the SEM image in Supplementary Fig.\  \ref{fig:buckling}.  To model nanobeam behavior in the presence of compressive stress and buckling, together with imperfect nanobeam shape and clamping points, we consider both an approximate analytic model, and finite elements simulations.  

The behavior of an ideal beam under axial compressive loads has been widely analyzed \cite{ref:saif2000tbm}.  For compressive axial load $F_i$, $\omega_m$ of a nanobeam with maximum deflection $\overline{x}$ can be expressed as
\begin{equation}\label{eq:wm_comp} 
\omega_m^2 = (\omega_m^i)^2\left(1- \frac{F_i}{F_c} +  3 \overline{x}^2 \frac{AE}{F_c} \frac{\pi^2}{(2L)^2} \right) 
\end{equation}
where $\omega_m^i$ is the resonance frequency of an ideal unloaded nanobeam, $A$ is the nanobeam cross--sectional area, $E$ is Young's modulus and $F_c$ is the critical buckling load. Equation \eqref{eq:wm_comp} is equivalent to the model presented in Ref.\ \cite{ref:kozinsky2006tnd}, and reveals the interplay between deflection, axial load, and stiffness.  For example, we see that ${d\omega_m}/{d\overline{x}}\propto  \overline{x}$, which is a result of more efficient conversion of transverse actuation to axial strain with increasing nanobeam curvature. Locally heating the nanobeam modifies $F_i \to F_i + \eta\epsilon  E\Theta$, where $\epsilon$ is the thermal expansion coefficient of the nanobeam and $0 < \eta \le 1$ is a geometric factor which accounts for non--uniform heating distribution of the nanobeam.  In general, $F_i$ and $\overline{x}$ are not independent.  However, in an ideally buckled nanobeam  \cite{ref:saif2000tbm}, we can show that 
\begin{align}\label{eq:dtdT}
\frac{d\overline{x}}{d\Theta} &=\frac{\mathcal{F}}{k}\frac{\kappa_t}{R}= \overline{x}\frac{\eta\epsilon w d E}{2 m \omega_m^2 L},
\end{align}
indicating that photothermal deflection can be enhanced in buckled nanobeams with $|\overline{x}| > 0$.  

Although the ideal nanobeam buckling model is instructive, it fails to reproduce experimental features such as deflection for axial load below the critical buckling load, and effects related to imperfect elastic clamping points and beam deformation \cite{ref:blocher2012add}.  To predict $\mathcal{F}$ and ${C}_t$ while including nanobeam non--idealities, we used finite element  ANSYS software to simulate $\omega_m$ and $\overline{x}$ as a function of axial and transverse loads, and for specified absorbed power. The results, summarized in Supplementary Fig.\ \ref{fig:preload} and Table \ref{tab:params}, indicate that significant enhancement of $\mathcal{F}$ compared to an unloaded nanobeam are expected. 

The simulations were conducted as follows.  The simulated structure consists of a nanobeam with dimensions $L\times w\times d = 80 \times 0.48 \times 0.25\,\mu\text{m}^3$, and includes the surrounding diamond chip.  Clamping point geometry  was found to significantly affect the simulated nanobeam properties. In fabricated nanobeams, the undercut process results in relatively complex clamping point geometry. Here the clamping points were modeled with a triangular vertical profile roughly approximating that of fabricated structures, extending $0.5\,\mu\text{m}$ on the nanobeam bottom surface and $1.0\,\mu\text{m}$ vertically along the undercut sidewall, as shown in Supplementary Fig.\ \ref{fig:preload}(a).  $\omega_m$ and $\overline{x}$ were then calculated as a function of $F_i$, as shown in Supplementary Fig.\ \ref{fig:preload}(b). Agreement between simulated and experimental $\omega_m$ was realized at two values of $F_i$, corresponding to pre-- and post--buckled nanobeam states, with $\overline{x} = -9\,\text{nm}$ and $-123\,\text{nm}$, respectively, where negative $\overline{x}$ indicates buckling down.

$\mathcal{F}$ and $C_t$ were estimated by simulating changes to $\overline{x}$ and $\omega_m$ when power $P_\text{abs}$ is uniformly absorbed across half of the nanobeam. During these simulations the bottom surface of the diamond substrate was fixed at constant temperature, and only conductive heat loss was considered. The corresponding temperature distribution was used to predict $R$ and $\kappa_t$, as summarized in Table \ref{tab:params}. The photothermal results are shown in Supplementary Fig.\ \ref{fig:preload}(c), which plots the changes $\Delta\overline{x}$ and $\Delta \omega_m$ in deflection and frequency, respectively, for $P_\text{abs} = 1\,\mu\text{W}$.  These results clearly illustrate  the sensitivity of photothermal effects on compressive stress and deflection, and indicate that $\mathcal{F}$ varies by over two orders of magnitude depending on the compressive axial loading of the nanobeam.

Simulations of $\Delta\omega_m$ when a vertical transverse load mimicking the optical gradient force is distributed along the coupling region of the nanobeam (length $L_c$) were also performed, and are shown in Supplementary Fig.\ \ref{fig:preload}(d).  These results show that $d\omega_m/dF_g$ varies by nearly three orders of magnitude depending on the compressive axial load.

\begin{figure}
\begin{center}
\epsfig{figure=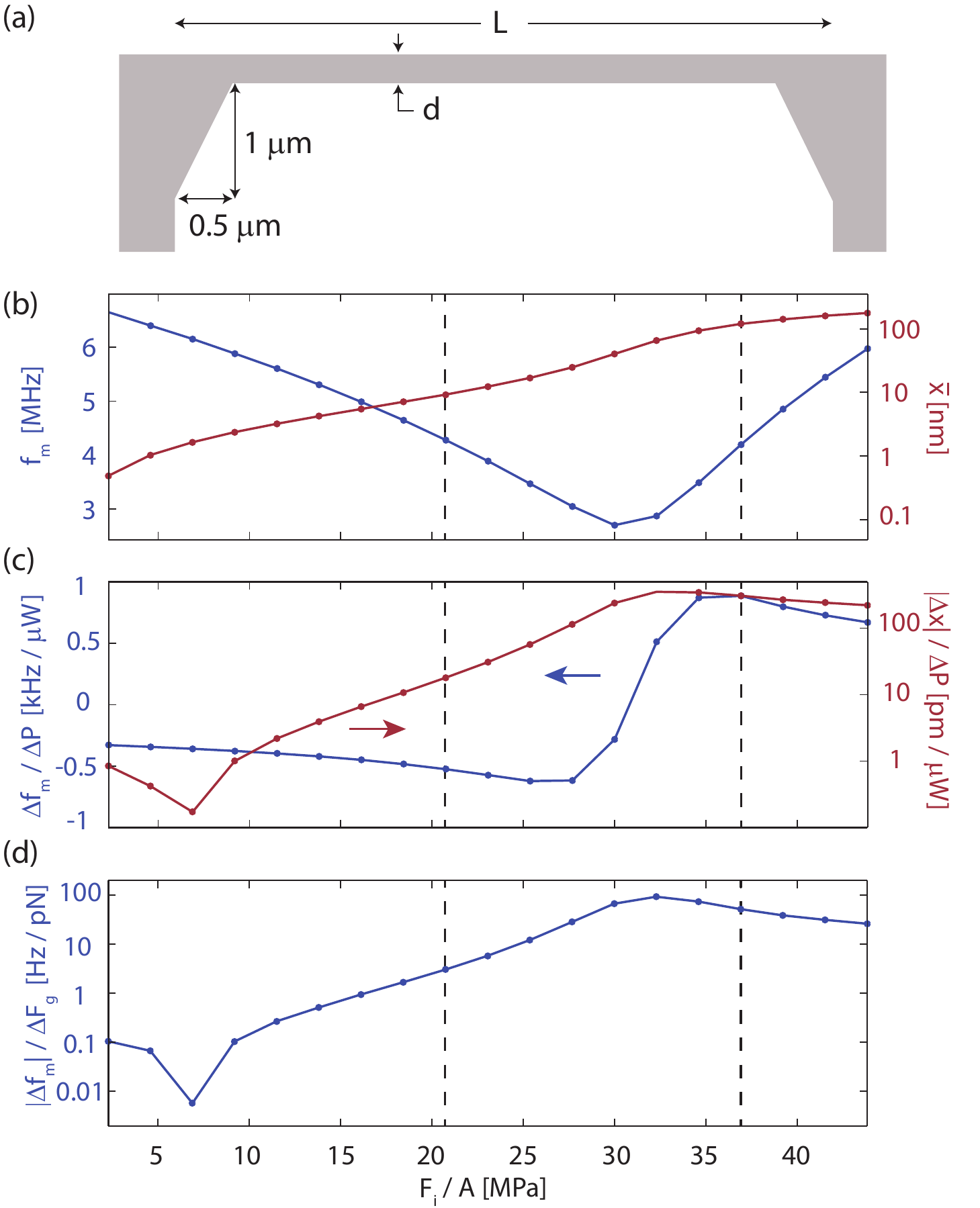, width=1\linewidth}
 \caption{(a) Cross--section of nanobeam used in ANSYS finite element simulations, showing the ``arched'' clamping point geometry. (b) Finite element simulations of resonance frequency and nanobeam deflection as a function of axial compressive stress.  Dashed lines indicate value of $F_i$ where simulated $\omega_m$ approximately matches the experimentally observed value. (c) Simulated change in resonance frequency and nanobeam deflection for $P_\text{abs} = 1\,\mu\text{W}$ of absorbed optical power. (d) Simulated change in resonance frequency from a $F_g = 1\,\text{pN}$ transverse load applied across the coupling region of the nanobeam (length $L_c$). Nanobeam dimensions are $L\times w\times d = 80 \times 0.48 \times 0.25\,\mu\text{m}^3$, as in the self--oscillation measurements.}
\label{fig:preload}
\end{center}
\end{figure}

\subsubsection{Dynamics: small--amplitude oscillations}

For small mechanical oscillation amplitude, the power in the nanobeam waveguide can be approximated by $P_n = P_n(h_o) - P_i\left.\frac{dT}{dx}\right|_{h_o} (x(t) - \overline{x})$. Inserting this into Eq.\ \eqref{eq:temp} and only retaining terms that are linear in $x$ allows the Laplace transform of  Eqs.\ \eqref{eq:eom} and \eqref{eq:temp}  to be combined into a single linear equation:
\begin{multline}
-\omega^2 x(\omega) -i\frac{\omega_m\omega}{Q_m} x(\omega) +\omega_m^2 x(\omega) = \\ \frac{F_{s}(\omega) }{m} +
\left(\frac{dF_{g}}{dx} + \frac{dF_{pt}}{dx}\frac{i\omega\tau +1}{\omega^2\tau^2+1}\right) \frac{ x(\omega)}{m} 
\end{multline}
where constant force terms resulting in changes to static nanobeam deflection $\overline{x}$ have been left out for clarity but are included implicitly in $\omega_m(\overline{x})$.  Rearranging terms reveals the optomechanical renormalization of the nanobeam dynamics,
\begin{equation}
-\omega^2 x(\omega) -i\gamma_m'\omega x(\omega) +\omega'^2_m x(\omega) = \frac{F_{s}(\omega) }{m} 
\end{equation}
with
\begin{align}
\frac{\omega_m'^2}{\omega_m^2} &= 1 - \frac{1}{1+\omega^2\tau^2}\frac{dF_{pt}(h_o)}{dx}\frac{1}{k} - \frac{dF_{g}(h_o)}{dx}\frac{1}{k}\label{eq:wm}\\
\frac{\gamma_m'}{\gamma_m } &=    1 + Q_m\frac{\omega_m\tau}{1+\omega^2\tau^2} \frac{dF_{pt}(h_o)}{dx}\frac{1}{k}\label{eq:gm},
\end{align}
where $dF_{pt}/dx = - \mathcal{F} \zeta L_i P_i\, dT/dx$.  Static effects modify the unperturbed resonance frequency $\omega_m^o$ according to
\begin{equation}\label{eq:wm_s}
\omega_m = \omega_m^o +  C_t \zeta L_i\,P_n({h})  + \frac{d\omega_m}{dF_g}F_g({h}).
\end{equation}
Equations \eqref{eq:wm}, \eqref{eq:gm} and \eqref{eq:wm_s} illustrate that combinations of mechanical softening or hardening, and amplification or damping are possible.  Higher order thermoelastic effects  \cite{ref:zaitsev2012ndm} are predicted to be small and are not included in this analysis.

\subsubsection{Nonlinear dynamics: large--amplitude oscillations}
When $\gamma_m'$ approaches zero, the amplitude of mechanical oscillations about the static (i.e., buckled) position grows, and nonlinear contributions to the system dynamics become significant.  These nonlinear modifications originate from mechanisms intrinsic to the nanomechanical device geometry, or to the optomechanical response of the system, and are characterized here by non--zero $\alpha_{2,3}$ and $d^{n}T/dx^{n}$ ($n>1$), respectively.  Two nonlinear features observed in Fig.\ 5 are a softening in $\omega_m$ at the onset of self--oscillation, and frequency harmonics in the self--oscillation region. While the later effect is significantly affected by the nonlinear response of the optomechanical system, the analysis of Zaitsev et al.\ \cite{ref:zaitsev2012ndm} shows that the optomechanical nonlinearity plays a negligible role in softening $\omega_m$ for the optomechanical system studied here. Rather, this softening is dominantly due to geometric nonlinearities of the deflected nanobeam.

Nonlinear coefficients $\alpha_2$ and $\alpha_3$ in Eq.\ \eqref{eq:eom}, can be derived from the  Euler--Bernoulli equation for a nanobeam with static deflection $\overline{x}$, as in Refs.\ \cite{ref:abou1993nrp,ref:lacarbonara1998evr}. Using an approximate ansatz for the static nanobeam shape, expressions for $\alpha_2$ and $\alpha_3$ can be derived \cite{ref:kozinsky2006tnd}.  For oscillations about $\overline{x}$, if we group the nanobeam deformations into time dependent and time independent parts according to
\begin{equation}\label{eq:deflected_beam_ansatz}
\phi(l,t) =\left(x(t)+\overline{x}\right)\frac{1}{2}\left(1 - \cos(2\pi l /L)\right), 
\end{equation}
where $l$ is the the coordinate running the length of the undeflected nanobeam, the  Euler--Bernoulli equation yields Eq.\ \eqref{eq:eom}, with nonlinear coefficients 
\begin{align}
\alpha_2 &= \overline{x}\frac{E}{6\rho}\left(\frac{2\pi}{L}\right)^4\frac{3}{8}\label{eq:alpha2},\\
\alpha_3 &= \frac{E}{18\rho}\left(\frac{2\pi}{L}\right)^4\frac{3}{8},\label{eq:alpha3}
\end{align}
where $\rho$ is the density of the nanobeam material. Note that while  $\alpha_2$ vanishes in a straight nanobeam, it is non--zero in deflected nanobeams ($|\overline{x}| > 0$).  

Assuming that the solution to Eq.\ \eqref{eq:eom} may be written as a combination of harmonic functions, the method of successive approximations \cite{ref:laudau1976mec, ref:abou1993nrp, ref:lacarbonara1998evr}  shows that to first order the fundamental frequency of oscillation is:  
\begin{align}
       \omega &=\omega_m(\overline{x}) + \frac{v^2}{\omega_m(\overline{x})}\left(\frac{3}{8}\alpha_3 - \frac{5 }{12\,\omega^2_m(\overline{x})}\alpha_2^2\right),\label{eq:freq_shift}\\
		&= \omega_m(\overline{x}) + v^2 \alpha\label{eq:freq_shift_tot},
\end{align}
where $v$ is the amplitude of oscillation and $\alpha$ is the effective nonlinear frequency shift coefficient. The first term in brackets is the well known Duffing frequency modification, whereas the second term results from nonlinearities induced by static deflection of the nanobeam.

\subsubsection{Parameter estimation and comparison with experiment}

\begin{figure}
\begin{center}
\epsfig{figure=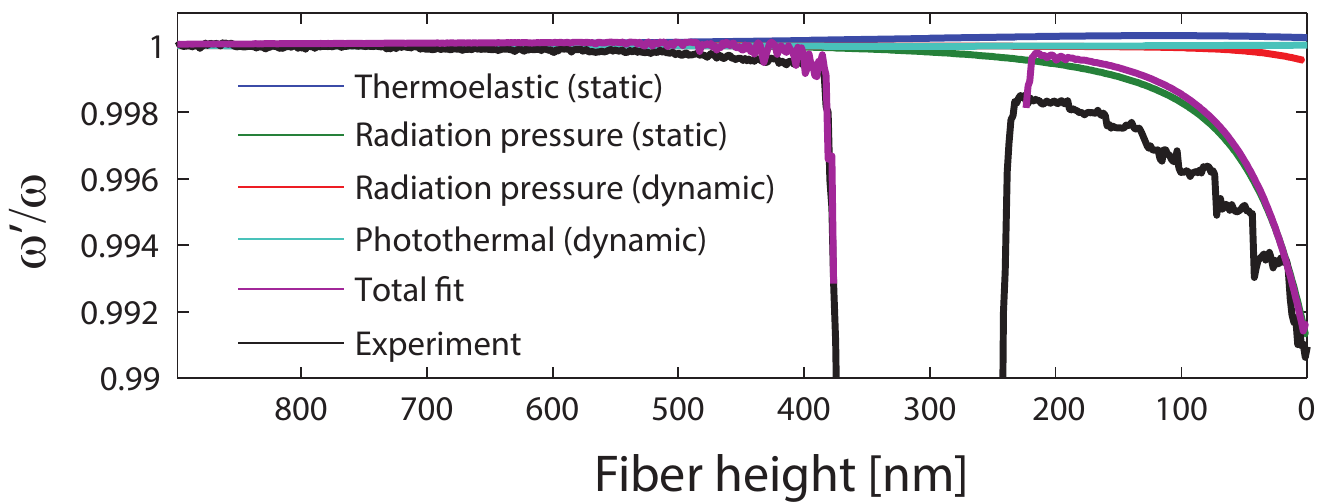, width=1\linewidth}
 \caption{Predicted $\omega'_m(h)$ generated using the model in Eq.\ \eqref{eq:wm} and the parameters in Table \ref{tab:params}.}
\label{fig:wm_comp}
\end{center}
\end{figure}

To compare the model described above with experimentally measured $\gamma_m'(h)$, as shown in Fig.\ 5(c), a combination of known, simulated and fit parameters were input into Eq.\ \eqref{eq:gm}, as summarized in Table \ref{tab:params} and described below.

For a given coupler operating condition, the photothermal force, and the resulting $\gamma'(h)$ described by Eq.\ \eqref{eq:gm}, scales linearly with $\zeta\mathcal{F}$. Neither $\mathcal{F}$ or $\zeta$ are  known \emph{a priori}. However, using finite element simulations to determine $\mathcal{F}$, as well as other parameters such as $\kappa_t$ and $R$, allows $\zeta$ to become the sole fitting parameter when comparing experimental and predicted values of $\gamma'(h)$. 

To determine $\mathcal{F}$ from finite element simulations, it is necessary to determine whether the nanobeam is in a pre-- or post--buckled state. Insight into the buckling configuration of the nanobeam is provided by the observed nonlinear softening at the onset of self oscillations, which is directly proportional to $v^2$  (Fig.\ 5(d)). From this data and Eq.\ \eqref{eq:freq_shift_tot}, the nonlinear coefficient $\alpha$ given in Table \ref{tab:params} can be measured. Together with Eqs.\ \eqref{eq:alpha2} and \eqref{eq:alpha3}, corresponding  nanobeam deflection amplitude of $|\overline{x}| = 98\,\text{nm}$ is inferred, in good agreement with $\overline{x} = -122\,\text{nm}$ predicted from finite element simulations of the nanobeam in its post--buckled configuration ($F_i/A \sim 37\,\text{MPa}$), as shown in Supplementary Fig.\ \ref{fig:preload}(b). When the nanobeam is in this post--buckled configuration, finite element simulations shown in Supplementary Fig.\ \ref{fig:preload}(c) indicate that $\mathcal{F} = -26\,\text{pN}/\mu\text{W}$ ($\mathcal{F}/k = -0.36\,\text{nm}/\mu\text{W}$).    Negative $\mathcal{F}$ and $\overline{x}$ indicate that the photothermal force and nanobeam deflection, respectively, are in the down direction.

\begin{table}[th]
\begin{tabular}{l r c l c}
Parameter 				& Input value 	  	& 			& Units							& Source \\  \hline\hline\\
$\kappa_t$ 				& $1.4$ 			&			&$ \,\mu \text{s}^{-1} $ 			& FEM simulation    \\
$R$						& $0.11 $ 		&			& K/$\mu$W$\mu\text{s}$ 			& FEM simulation \\
$\omega_m^i/2\pi$	      	& 680 			&			& kHz					      		& FEM simulation  \\
$m$						& 10  			&			& pg							      	& FEM simulation  \\
$\overline{x}	$			& $-122$ 	 	&			& nm							& FEM simulation  \\
$\alpha/2\pi $				& $-16\, (-28)$		&		& $\text{Hz}/\text{nm}^2$			& Fit (FEM) \\
$C_t$ 					& $1.0$			&  			& kHz/$\mu$W					& FEM simulation \\
$\zeta$					& 0.12			&			& $\text{cm}^{-1}$				& Fit  \\
$\mathcal{F}/k$ 		      	& $-0.36$		& 			& nm/$\mu$W				      	& FEM simulation \\
$\mathcal{F}$ 		      	& $-26$			& 			& pN/$\mu$W				      	& FEM simulation \\
$F_g $ 					& Sup.\ Fig.\ \ref{fig:fg}&    		& 								& Optical mode solver  \\
$d\omega_m/dF_g$	      & $-7163$ ($-377$)	& 			& rad/s\, pN$^{-1}$       		& Fit (FEM) \\
$P_i$					& 300			&			& $\mu$W						& Experimental parameter \\
$Q_m$					& 25000			&			& 								& Experimental parameter \\
\end{tabular}
\caption{Parameters input to model for $\gamma'_m(h)$ and $\omega'_m(h)$. Values in brackets are predictions from simulations, and are included for comparison with values determined from experimental fits.}\label{tab:params}
\end{table}

Inputing $\mathcal{F}$ and other finite element simulated parameters summarized in Table \ref{tab:params} into Eq.\ \eqref{eq:gm}, good agreement between predicted and observed $\gamma_m'(h)$ was found for $\zeta \sim 0.12\,\text{cm}^{-1}$. The corresponding optical absorption rate can be described by quality factor $Q_o = 2\pi\,n_g/\zeta\lambda \sim 6.6 \times 10^5$, where the group index $n_g \sim 2.0$ of the nanobeam is predicted from numerical simulation.  This absorption rate is smaller than combined absorption and radiation loss rates in other single crystal diamond nanophotonic structures \cite{ref:burek2014hqf}.

Given  the value of $\zeta$ obtained from fitting $\gamma'(h)$,  $\omega'_m(h)$ predicted from Eq.\ \eqref{eq:wm} can be compared with measurements. This is shown in Supplementary Fig.\ \ref{fig:wm_comp}, in which the predicted $\omega_m'(h)$ was generated with  $d\omega_m/dF_g$  as a fitting parameter, and other parameters set as in the model for $\gamma'(h)$ and listed in Table \ref{tab:params}. Static and dynamic thermal effects are found to be significantly smaller than the maximum experimentally observed shift to $\omega_m$.  Rather, the monotonic decrease in $\omega_m'$, which becomes significant for $h< 200\,\text{nm}$, follows an $h$ dependence consistent with static tuning by the attractive optical gradient force $F_g(h)$ pulling up on a down--buckled nanobeam. This is in contrast to the static thermal tuning described by $C_t$, which is proportional to the power coupled into the nanobeam, and is expected to decrease in magnitude with decreasing $h$ for $h < 200\,\text{nm}$.  

There is good agreement between the model and experimental observation in Supplementary Fig.\ \ref{fig:wm_comp},  however the fit value for $d\omega_m/dF_g$ is larger than expected from finite element simulations (Supplementary Fig.\ \ref{fig:preload}(d)) and predictions of $F_g(h)$ (Supplementary Fig.\ \ref{fig:fg}).  As shown in Supplementary Fig.\ \ref{fig:preload}(d), $d\omega_m/dF_g$ was found in simulations to be highly variable, spanning over three orders of magnitude, depending on compressive stress and resulting buckling configuration. It is possible that imperfect nanobeam and clamping point shape result in an enhanced sensitivity \cite{ref:lacarbonara1998evr, ref:ouakad2010dbm}.  For example, including the triangular ``arched'' clamping points to approximate the fabricated structure, as shown in Supplementary Fig.\ \ref{fig:preload}(a),  enhanced $d\omega_m/dF_g$ by a factor of $\sim 2$ compared to the case of ideal clamping points.  Other effects not included here include reflection of light from the end of the waveguide resulting in enhance optical interactions, dynamic and static displacement of the fiber taper due to optical forces \cite{ref:eichenfield2007ams}, low frequency fiber vibrations and possible parametric driving for small $h$, breakdown of the coupler two mode representation and the influence of higher--order modes for small $h$, and short range effects such as the Casmir force \cite{ref:pernice2010asr}.  

In conclusion, the analysis presented here serves to illustrate the influence of compressive loading and buckling on the photothermal response of nanobeams, and to show that significant photothermal forces are present at relatively low optical absorption levels. Note that imperfect nanobeams have significantly altered pre-- and  post-- buckling behavior when subject to an compressive load \cite{ref:fang1994pbm}, and taking into account these non--idealities is necessary to more accurately predict the nanobeam behavior.

%\bibliography{nano_bib}


\end{document}